\documentclass{article}

\usepackage{arxiv}

\usepackage[utf8]{inputenc} % allow utf-8 input
\usepackage[T1]{fontenc}    % use 8-bit T1 fonts
\usepackage{hyperref}       % hyperlinks
\usepackage{url}            % simple URL typesetting
\usepackage{booktabs}       % professional-quality tables
\usepackage{amsfonts}       % blackboard math symbols
\usepackage{nicefrac}       % compact symbols for 1/2, etc.
\usepackage{microtype}      % microtypography
\usepackage{lipsum}		% Can be removed after putting your text content
\usepackage{graphicx}
\usepackage{natbib}
\usepackage{amsthm}
\usepackage{amsmath}
\usepackage{doi}

\title{Are we misdiagnosing ensemble forecast reliability? \\ On the insufficiency of Spread-Error and rank-based reliability metrics}

\author{
    \textbf{Arlan Dirkson} \quad 
    \textbf{Mark Buehner} \\[0.5em]
    \normalsize Data Assimilation and Satellite Meteorology \\ 
    \normalsize Environment and Climate Change Canada \\ 
    \normalsize Quebec, Canada \\[0.8em]
    \normalsize Corresponding author: Arlan Dirkson \\ 
    \normalsize \texttt{arlan.dirkson@ec.gc.ca}
}
\date{}

\renewcommand{\shorttitle}{Insufficiency of Spread-Error and rank-based reliability metrics}

\begin{document}
\maketitle

\begin{abstract}
It has been documented that Spread-Error equality and a flat rank histogram are necessary but insufficient for demonstrating ensemble forecast reliability. Nevertheless, these metrics are heavily relied upon---both in the literature and at operational numerical weather prediction centers---as if they are valid indicators of perfect ensemble dispersion. In this study, we demonstrate theoretically why the Spread-Error relationship is necessary but insufficient for diagnosing reliability up to second order, even when unconditional bias is absent or accounted for. Assuming joint normality between ensemble members and the reference truth, we further show with idealized experiments that the same covariance structure responsible for this insufficiency also produces false diagnoses of reliability with the rank histogram and the reliability component of the continuous rank probability score. Under this structure and when the ensemble-mean state is meaningfully different from climatology, the truth lies among the least extreme members when climatological variance is excessive in each member, and consistently among the most extreme members when climatological variance is deficient. Importantly, this behavior is also shown to be plausible in operational ensemble weather forecasts. Combining these results with calibration principles from statistical postprocessing leads us to conclude that both ``perfect dispersion'' and ``underdispersion'' are ill-defined. When diagnostics are misinterpreted as indicating the latter, improper tuning of forecasts can lead to further deterioration of forecast quality, even while improving Spread-Error and rank histogram behavior. To address these issues, we propose a new reliability diagnostic based on three easily computed statistics, motivated by the structure of the joint distribution of ensemble members and the reference truth up to second order. The diagnostic separates contributions to unreliability originating from climatology and from predictability, enabling a more precise and robust characterization of ensemble behavior.
\end{abstract}

% keywords can be removed

\defcitealias{wilks2011reliability}{W11a}
\defcitealias{marzban2011effect}{M11}
\defcitealias{rodwell2016reliability}{R16}
\defcitealias{dirkson2025impact}{D25}

\newtheorem{proposition}{Proposition}

\section{Introduction}\label{sec:intro}
It has been proposed that the fundamental goal of an ensemble forecast is to maximize \textit{sharpness} while maintaining \textit{reliability} \citep[][]{gneiting2007probabilistic}. Maximizing sharpness refers to minimizing the ensemble spread (i.e., its standard deviation or variance), whereas reliability---also termed \textit{calibration} or \textit{ensemble consistency}---pertains to the accurate representation of forecast uncertainty \citep[e.g.][]{palmer2000predicting,mason2008we}. Reliability takes on slightly different forms in the ensemble forecasting literature. In some contexts, it refers to the statistical indistinguishability between the ensemble members and the true state, typically assessed by the flatness of a rank histogram \citep{anderson1996method,talagrand1997evaluation,hamill1997rank} or by comparing the mean squared error (MSE) of the ensemble mean and the average ensemble variance \citep[e.g.][]{murphy1988impact,leutbecher2008ensemble}. This latter diagnostic is known as the Spread-Error relationship, or, if square roots are taken, root mean square (RMS) error versus RMS spread. In other contexts, reliability refers to the property that forecast probabilities for meteorological events (e.g., the probability that the maximum temperature exceeds 35$^\circ$C) match their observed frequencies when conditioned on the forecast probabilities themselves. The reliability of such conditional probabilities is evaluated using a reliability diagram \citep{murphy1977reliability,hamill1997reliability,stephenson2003forecast}.

Both forms of reliability derive from a classic definition that \cite{gneiting2007probabilistic} termed {\it complete calibration}. For a forecast probability distribution $f_t$ at time $t$, an $n$-member ensemble, $x_{1,t}, \dotsc, x_{n,t}$, drawn from $f_t$ is completely calibrated if the true state, $y_t$, is drawn from a distribution $g_t$ satisfying $g_t=f_t$ for all $t=1,\dotsc$. While theoretically useful, complete calibration is not practically testable since $g_t$ cannot be inferred from a single realization $y_t$. In practice, the statistical consistency between ensemble members and the true state, or the reliability of forecast probabilities, must be evaluated in an aggregate sense over a collection of ensemble forecasts and reference truths.

\cite{wilks2011reliability} (hereafter \citetalias{wilks2011reliability}) emphasized that when considering a collection of ensemble forecasts in the aggregate, reliability pertains to the joint probability distribution linking ensemble members and the true state. This distribution captures both climatology---through marginal distributions---and predictability---through conditional distributions. Climatology in this context refers to the distribution of possible states that can occur over a validation period (e.g., at a particular location and lead time). Differences between forecast climatology and true climatology are affected by both the underlying climatology of the model (e.g., in a freely-running model or at very long lead times) as well as the assimilation processes which updates forecasts with new observations. Predictability refers to both that within the ensemble itself---also known as potential predictability---as well as that relative to the reference truth---also known as actual or true predictability \citep[e.g.,][]{anderson1996evaluating,delsole2005predictability,kumar2014there}. The relevance of the joint distribution in the context of reliability is also broadly understood in the field of statistical postprocessing \citep[e.g.,][]{li2017review}. 

\cite{broecker2011concept} explicitly framed reliability through the lens of the joint distribution in terms of {\it exchangeability}, a property from multivariate statistics also cited in \citetalias{wilks2011reliability}. In this context, exchangeability means that the joint distribution for the sequence of random variables $Y,X_1,\dotsc,X_n$ (with $Y$ as the true state and $X_1,\dotsc,X_n$ the ensemble members) is invariant under any permutation of the sequence. Exchangeability implies that the marginal/climatological distribution of each ensemble member equals that of the truth and that potential predictability within the ensemble matches actual predictability relative to the truth \citepalias{wilks2011reliability}. \cite{broecker2011concept} showed theoretically how exchangeability results in flat rank histograms, as expected for reliable ensembles. This was also demonstrated in \cite{marzban2011effect} (hereafter \citetalias{marzban2011effect}) with idealized experiments based on the multivariate normal distribution. \citetalias{wilks2011reliability} applied the same idealized system and provided evidence that exchangeability also ensures correct event probabilities in reliability diagrams, aside from expected deviations due to small ensemble size \citep{richardson2001measures}. \cite{fortin2014should} and \cite{dirkson2025impact} (hereafter \citetalias{dirkson2025impact}) offered independent proofs that the Spread-Error relationship also holds under exchangeability. 

Studies have demonstrated however that a flat rank histogram is necessary but insufficient for demonstrating reliability \citep[e.g.,][]{hamill2001interpretation,gneiting2007probabilistic,bishop2008bayesian}.  \cite{hamill2001interpretation} showed that uniformity can result from distributions $f_t$ and $g_t$ that, at three different times, are not equal---two being subject to opposing conditional biases and one being subject to excessive dispersion. \cite{bishop2008bayesian} showed that a flat rank histogram can also be produced when forecasts are subject to excess climatological variance. To detect false diagnoses of reliability with the rank histogram, \cite{gneiting2007probabilistic} (and later \citetalias{marzban2011effect}) recommended separately evaluating {\it marginal calibration}---i.e., equivalence between the climatological distributions describing the forecasts and the reference truth---using quantile-quantile plots. With the basic idea being that the rank histogram can only assess reliability in an unconditional sense, others have proposed stratifying forecasts based on different forecast scenarios and verifying that each subset produces a flat rank histogram \citep{hamill2001interpretation,brocker2008reliability,siegert2012rank,brocker2020stratified}.

Under a similar theme, \cite{johnson2009reliability} claimed that the Spread-Error relationship is necessary but insufficient for diagnosing reliability (even up to second order). They posited that to do so, one must also show that forecasts are unbiased and exhibit a climatological variance that is consistent with that of the truth. Verifying these two conditions is equivalent to verifying marginal calibration up to second order. This claim was presented in the context of showing that a statistical postprocessing technique---known as member-by-member calibration---satisfies all three conditions, but was otherwise lacking details on why Spread-Error equality is insufficient. A more common (albeit still rare) acknowledgment is that mean bias complicates Spread-Error comparison, since its contribution to the ``Error'' term mimics underdispersion \citep[\citetalias{wilks2011reliability},][]{goddard2013verification,bauer2016aspects,rodwell2016reliability,wang2018sensitivity}. This in part motivated the development of the Reliability Budget of \cite{rodwell2016reliability} (hereafter \citetalias{rodwell2016reliability}), which explicitly removes the contribution of mean bias from the original Spread-Error diagnostic. Alternatively, mean bias can simply be removed from forecasts prior to evaluating Spread-Error consistency \citep[e.g.,][]{goddard2013verification}. 

Although the limitations of the rank histogram and the Spread-Error relationship are partially documented, it remains common practice to restrict reliability assessments to these diagnostics, without accounting for confounding effects. For instance, ensemble tuning and routine monitoring of reliability at operational forecasting centers often rely exclusively on the Spread-Error relationship, particularly in ensemble data assimilation and medium-range forecasting \citep[e.g., ][]{bonavita2012use,berner2017stochastic,haiden2018,buehner2020local,mctaggart2022using,inverarity2023met}. As we will show, such reliance can lead to adjustments that inadvertently degrade forecast quality, which has the potential to misdirect research efforts when improvements to reliability are pursued solely with respect to these diagnostics. 

In this study, we investigate various reliability metrics with respect to their ability to detect ensemble unreliability. This is done by framing reliability in terms of the exchangeability property, following \citetalias{wilks2011reliability} and \cite{broecker2011concept}. We show that in the presence of a climatological variance bias, several reliability diagnostics can falsely indicate that forecasts are perfectly reliable. These include the Spread-Error relationship, Reliability Budget, and, assuming joint-normality, the rank histogram and reliability component of the continuous rank probability score (CRPS) \citep{hersbach2000decomposition}. Combining theoretical results for the Spread-Error relationship with calibration principles from statistical postprocessing leads us to conclude that both ``perfect dispersion'' and ``underdispersion'' are ill-defined through Spread-Error and rank-based framings. To address these issues, we propose a simple diagnostic procedure that separates climatological unreliability from unreliability related to predictability. Up to second order, this distinction is made using three simple statistics computed at each location and lead time over a validation set: climatological mean bias, climatological variance bias, and a linear predictability bias (via correlations).

The remainder of this study is organized as follows. Section \ref{sec:metrics} describes the reliability diagnostics that will be considered. The statistical framework for reliability is discussed in Section \ref{sec:framework}. Theoretical results in Section \ref{sec:pitfalls} prove that the Spread-Error relationship and Reliability Budget are necessary but insufficient second-order diagnostics. Idealized experiments are conducted in Section \ref{sec:idealized} to illustrate these theoretical results and place them into broader context with other reliability diagnostics. A real-world example of this insufficiency in an operational ensemble system is shown in Section \ref{sec:real_forecasts}. In Section \ref{sec:qualifying}, the statistical framework is extended and combined with calibration principles to explain why framing reliability in terms of ensemble spread can be misleading. An alternative diagnostic framework for evaluating reliability up to second order is described in Section \ref{sec:new_diagnostic}. A summary is provided in Section \ref{sec:summary}.

\section{Reliability Diagnostics}\label{sec:metrics}
\subsection{Spread-Error Relationship}
The variance of a single ensemble forecast is a natural measure of forecast uncertainty \citep[e.g.][]{leith1974theoretical}. The unbiased ensemble variance at validation time $t$ at a particular location and lead time is defined:
\begin{equation}\label{eq:ens_spread}
    S_{e_t}^2 = \frac{1}{n-1} \sum_{i=1}^{n} (x_{i,t}-\overline{x}_t)^2,
\end{equation}
where $x_{i,t}$ denotes the $i$th ensemble member, $\overline{x}_t=\frac{1}{n}\sum_{i=1}^{n} x_{i,t}$ is the ensemble mean, and $n$ is the ensemble size. Its average over $T$ validation times is: 
\begin{equation}\label{eq:ens_spread_average}
    \langle S_{e_t}^2 \rangle = \frac{1}{T} \sum_{t=1}^{T} S_{e_t}^2,
\end{equation}
where the angled brackets, $\langle \cdot \rangle$, denote a time average.

The MSE (Mean Squared Error) of the forecast ensemble mean relative to the true state $y_t$ is defined:
\begin{equation}\label{eq:mse_tau}
    \mathrm{MSE}_\tau = \left\langle \left(\overline{x}_t - y_t\right)^2\right\rangle,
\end{equation}
where subscript $\tau$ emphasizes that this definition is with respect to the true state (as opposed to observations). 

When ensemble members are exchangeable with the truth, it can be shown that Eqs. \ref{eq:ens_spread_average} and \ref{eq:mse_tau} are related in expectation by \citep[][\citetalias{dirkson2025impact}]{fortin2014should}:
\begin{equation}\label{eq:spread_error}
    \mathbb{E}[\mathrm{MSE}_\tau] = \frac{n+1}{n} \mathbb{E}[\langle S_{e_t}^2 \rangle].
\end{equation}
This expression was also presented under the more restrictive assumption of independence in \cite{leutbecher2008ensemble} and under unknown assumptions in \cite{murphy1988impact}. In practice, the Spread-Error relationship is assessed outside of expectation (i.e., based on averages) by comparing ``Spread'' and ``Error'' either side-by-side \citep[e.g.][]{houtekamer1998data,hagedorn2008probabilistic}, as a scatter plot based on binned values \citep[e.g.][]{bonavita2012use,haiden2018} (also sometimes normalized by the climatological variance), or by constructing a scalar metric based on their ratio or difference \citep[e.g.][]{ho2013examining}. 

Here, the Spread-Error difference as defined in \citetalias{dirkson2025impact} is considered:
\begin{equation}\label{eq:delta_tau}
    \delta_\tau = \frac{n+1}{n}\langle S_{e_t}^2 \rangle - \mathrm{MSE}_\tau.
\end{equation}
The quantity $\delta_\tau$ is unbiased with respect to ensemble size $n$ and sample size $T$. Under common interpretation, $\delta_\tau>0$ is associated with overdispersion, $\delta_\tau<0$ is associated with underdispersion, whereas $\delta_\tau=0$ is associated with perfect dispersion. These interpretations come with the important (though often unrecognized) caveat that mean bias contributes negligibly to the ``Error'' component \citep[e.g., \citetalias{wilks2011reliability},][]{goddard2013verification,rodwell2016reliability,bauer2016aspects,wang2018sensitivity}, which is rare in real systems.

When evaluated against observations, the MSE in the Spread-Error difference can be modified to account for observation error through \citep[e.g.,][\citetalias{rodwell2016reliability}]{bowler2008accounting}:
\begin{equation}\label{eq:delta_tau_hat}
    \delta_{\hat{\tau}} =  \frac{n+1}{n}\langle S_{e_t}^2 \rangle - \left(\mathrm{MSE}_o - \sigma_o^2\right),
\end{equation}
where $\mathrm{MSE}_o$ is the MSE against observations and $\sigma_o^2$ is the observation error variance. Equations \ref{eq:delta_tau} and \ref{eq:delta_tau_hat} are equal in expectation under typical assumptions that observations are unbiased and that their error is independent from the ensemble forecasts and true state.

\subsection{Reliability Budget}
\citetalias{rodwell2016reliability} adapted Eq. \ref{eq:delta_tau_hat} via their Reliability Budget to account for mean forecast bias, which, as noted above, complicates interpretation. That budget (given by their Eq. 4) can be written: 
\begin{equation}\label{eq:reliability_budget}
    \begin{aligned}
        \delta_{\hat{\tau}}^\nu &= \frac{n+1}{n}\langle S_{e_t}^2\rangle \hspace{5em}~~~~ \text{EnsVar} \\ &~~- \frac{T}{T-1} \mathrm{MSE}_o \hspace{5em} \text{Depar}^2 \\
         &~~+ \frac{T}{T-1}\langle \overline{x}_{t} - y_{o,t} \rangle^2 \hspace{2.9em} \text{Bias}^2 \\  &~~+ \sigma_o^2. \hspace{9em}\text{ObsUnc}^2
    \end{aligned}
\end{equation}
The quantity $\delta_{\hat{\tau}}^\nu$ on the left-hand-side of Eq. \ref{eq:reliability_budget} is equivalent to the ``residual'' term in their Eq. 4, but multiplied by $-1$ for consistency with Eq. \ref{eq:delta_tau_hat}. The superscript $\nu$ signifies that $\delta_{\hat{\tau}}^\nu$ is intended to be a measure of reliability related specifically to the variance (or spread) of the ensemble, isolated from the effects of mean bias that are removed through the subtraction of the $\text{Bias}^2$ term. Importantly, this interpretation assumes that mean bias is stationary, which is why the budget must be computed at individual locations before performing any spatial averaging. \citetalias{rodwell2016reliability} acknowledged that mean bias must still be assessed separately as part of the reliability evaluation. When Eq. \ref{eq:reliability_budget} is expressed instead in terms of the true state (as will be done later in idealized experiments), $\sigma_o^2$ is set to zero and $\delta_{\hat{\tau}}^\nu$ is denoted $\delta_{\tau}^\nu$. Both $\delta_{\hat{\tau}}^\nu$ and $\delta_{\tau}^\nu$ will be referred to simply as the ``Reliability Budget'' throughout this study. 

\subsection{Rank Histogram}
The rank histogram \citep{anderson1996method,talagrand1997evaluation,hamill1997rank} is formulated by ordering the members of an ensemble forecast at time $t$ from smallest to largest, and counting the rank of the reference truth among the ordered ensemble. The collection of ranks over $T$ forecasts are then plotted as a histogram distribution with $n+1$ bins. Different histogram shapes are typically used to infer bias (sloped distribution), underdispersion (U-shaped distribution), and overdispersion (dome-shaped distribution). We quantify departures from uniformity using the $\chi^2$ test \citep{jolliffe2008evaluating}.

\subsection{Reliability Diagram}    
The reliability diagram \citep[e.g.,][]{murphy1977reliability,hamill1997reliability,stephenson2003forecast} plots the observed frequency of a categorical event (e.g., that rainfall exceeds 10 mm), denoted $\overline{o}_k$, given a forecast probability of the event $p_k=\frac{k}{n}$ ($k = 0, 1, \dots, n$) is issued. The $k$th observed frequency is computed as the fraction of times the event occurs, conditioned on the forecast probability $p_k$ being issued. When $\overline{o}_k$ and  $p_k$ are plotted against each other, reliable forecasts should result in points (often connected by a line) following the 1:1 line. However, even if ensemble members are consistent with the truth, this may not result in reliable forecast probabilities unless the ensemble size is sufficiently large \citep[][\citetalias{wilks2011reliability}]{richardson2001measures}. Smaller ensemble sizes tend to tilt reliability curves toward a shape indicative of overconfidence (reversed S-shaped), particularly when predictability is low. Other shapes are used to infer bias (consistent over- or under-prediction) and underconfidence (S-shaped) \cite[see e.g.,][]{wilks2011statistical}. We quantify departures from the 1:1 line using the reliability component of the Brier Score decomposition \citep{murphy1973new}, denoted $\mathrm{BS_{rel}}$. This scalar quantity is minimized at zero, corresponding to when the reliability curve follows the 1:1 line.

\subsection{CRPS reliability}
The continuous rank probability score (CRPS) is given by:
\begin{equation}\label{eq:crps}
    \mathrm{CRPS} =  \int \left(\tilde{F}_t(x) - H(x-y_t)\right)^2 d x,
\end{equation}
where $\tilde{F}_t$ denotes empirical cumulative distribution function (CDF) of the ensemble forecast at time $t$, and $H(x)$ is the Heaviside function used to define the CDF for the truth (a step function at $y_t$). Equation \ref{eq:crps} is then averaged over $T$ forecasts. Like the Brier score, the mean CRPS can be decomposed to yield a reliability component \citep[see details in][]{hersbach2000decomposition}, which we denote by $\mathrm{CRPS_{rel}}$. $\mathrm{CRPS_{rel}}$ measures something similar to the rank histogram, but takes into account the average width of each bin (distance between ordered members). Like $\mathrm{BS_{rel}}$, $\mathrm{CRPS_{rel}}$ is also optimized at a value of zero.

\section{Statistical Framework}\label{sec:framework}
\begin{figure*}[t]
  \centering
  \noindent\includegraphics[width=1.0\textwidth,angle=0]{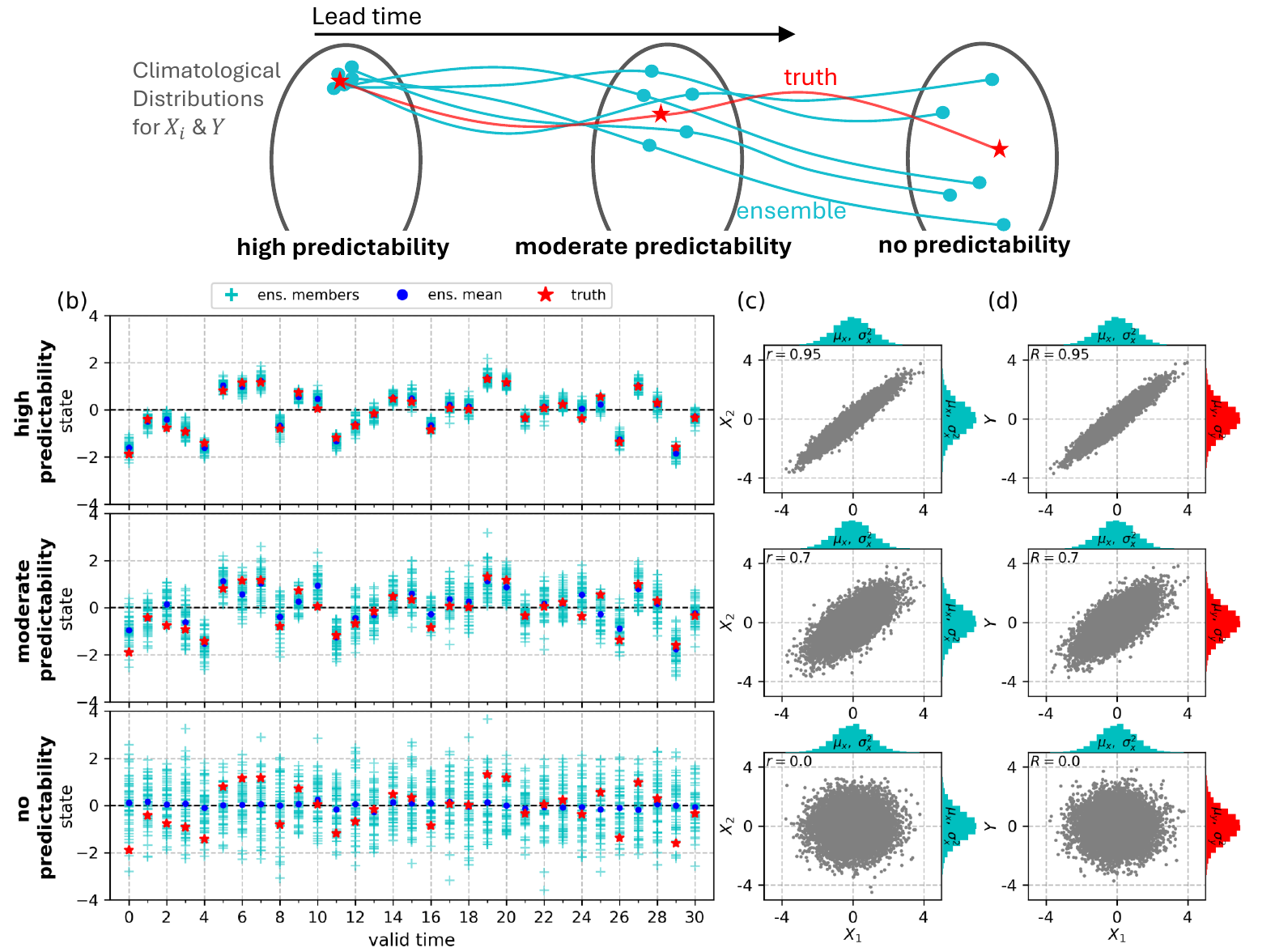}\\
  \caption{(a) Schematic of a 5-member ensemble forecast and reference true state as a function of lead time, transitioning through high, moderate, and no predictability; gray ellipses represent the climatological distributions of possible states, drawn to be the same for $X_i$ and $Y$. (b) Time series of 50-member ensemble forecasts and true states simulated with the multivariate normal distribution under each predictability regime highlighted in (a). (c) Scatter plots of the first and second ensemble member. (d) Scatter plots of the first ensemble member and the true state. The cyan histograms along the axes in (c) and (d) illustrate the climatological distribution for the corresponding ensemble member, whereas the red histograms in (d) illustrate the climatological distribution for the true state. The forecasts displayed in (b)-(d) are reliable, as they satisfy the three conditions for exchangeability up to second order described in the main text.}\label{f1}
\end{figure*}
The following description closely follows that of \citetalias{wilks2011reliability} and \citetalias{dirkson2025impact}. Let $X_1, \dotsc, X_n$ denote a sequence of random variables representing $n$-total ensemble members, and let $Y$ be a random variable for the true state. These variables are to represent the ensemble and truth at a particular location and lead time. The sequence of random variables $Y,X_1,\dotsc,X_n$ follows a joint probability distribution, emphasizing the fact that these variables are not independent over a collection of validation times when there is predictability in the system. The joint distribution is simplified by assuming ensemble members are exchangeable with one another. Up to second order, exchangeability implies that each ensemble member shares the same expected value, $\mathbb{E}[X_i] = \mu_x$, and variance, $\mathbb{V}(X_i) = \sigma_x^2$, where $\mu_x$ and $\sigma_x^2$ represent the mean and variance of the marginal (i.e., climatological) distribution for each member. Exchangeability furthermore implies that the inter-member covariance, $\mathrm{Cov}(X_i, X_j)$, is constant for all $i\neq j$. The first two moments of the marginal distribution for the truth are similarly $\mathbb{E}[Y]=\mu_y$ and $\mathbb{V}[Y]=\sigma_y^2$. The member-truth covariance is denoted $\mathrm{Cov}(X_i, Y)$, which is a constant for all $i=1,\dotsc,n$ due to exchangeability between members. Finally, it is useful to define the inter-member correlation, $r = \mathrm{Cov}(X_i, X_j) / \sigma_x^2$, which is a measure of potential predictability. The member-truth correlation $R = (\sigma_x \sigma_y)^{-1} \mathrm{Cov}(X_i, Y)$ is similarly a measure of actual predictability.

\citetalias{wilks2011reliability} and \cite{broecker2011concept} proposed that reliability is achieved when ensemble members are not only exchangeable with each other, but when they are also exchangeable with the true state. Members $X_1,\dotsc,X_n$ are exchangeable with $Y$ up to second order if the following three conditions hold [see proposition 3.1 in \cite{niepert2014exchangeable} as well as \citetalias{wilks2011reliability} and \citetalias{dirkson2025impact}]:
\begin{enumerate}
\item climatological mean condition: $\mu_x=\mu_y$
\item climatological variance condition: $\sigma_x^2=\sigma_y^2$
\item covariance condition: $\mathrm{Cov}(X_i,X_j)=\mathrm{Cov}(X_i,Y)$ \\ {\it or} \\ linear predictability condition: $r=R$.
\end{enumerate}
Reliability conditions 1 and 2 pertain to marginal calibration (i.e., climatological reliability), and require that the mean and variance of the climatological distribution for each member match those of the truth. The variances and covariances (or correlations) in reliability conditions 2 and 3 form the covariance matrix of the joint distribution. Reliability condition 3 depends on the conditional relationships between ensemble members and between ensemble members and the truth. It was expressed in terms of covariances in \cite{niepert2014exchangeable} and \citetalias{dirkson2025impact}, whereas \citetalias{wilks2011reliability} considered the condition in terms of the correlations $r$ and $R$. 

If the joint distribution is non-stationary, reliability conditions 1--3 should be written explicitly as functions of time, $t$. Soon, we consider these conditions expressed as differences ($\mu_x - \mu_y$, $\sigma_x^2 - \sigma_y^2$, etc.), which only need to vary with time if the differences themselves are non-stationary. In this study, we make the simplifying assumption that these differences are stationary. Methods for addressing violations of this assumption related to the mean state were discussed in \cite{goddard2013verification}.
 
To build further intuition for the joint distribution, the evolution of a single ensemble forecast with $n=5$ members is depicted in Fig. \ref{f1}a as a function of lead time. Three lead times are highlighted corresponding to high, medium, and no predictability. The climatological distributions for the members and truth are depicted as being equal, indicative of climatological reliability. For simplicity, they are also depicted as being equal for all three lead times. The amount of predictability can be qualitatively inferred by the phase space occupied by the ensemble members and true state relative to that occupied by their climatological distributions. At each lead time, the ensemble and true state are described by a different joint distribution.

To illustrate the joint distribution at each lead time over a collection of forecasts, Fig. \ref{f1}b-d shows samples of ensemble forecasts and true states simulated with a multivariate normal (MVN) distribution. This idealized system, which was also used in \citetalias{marzban2011effect} and \citetalias{wilks2011reliability} to study the rank histogram and reliability diagram, will be used throughout this study. Here, the climatological distributions for the ensemble and true state are assigned to the standard normal distribution (i.e., $\mu_x=\mu_y=0$ and $\sigma_x^2=\sigma_y^2=1$). The level of potential predictability is controlled by setting the inter-member correlation at the three respective lead times to $r=0.95$, $r=0.6$, and $r=0$, and actual predictability is (in this case) set to match potential predictability. As such, all three conditions for reliability up to second order described previously are satisfied. 

At the shortest lead time, time series in Fig. \ref{f1}b (top row) show ensemble members and the true state clustering tightly around the varying ensemble mean, consistent with high sharpness and strong predictability ($r=R=0.95$). This is also evident in the scatter plots between $X_1$ and $X_2$ in Fig. \ref{f1}c and between $X_1$ and $Y$ in Fig. \ref{f1}d (top row), where the points closely following the 1:1 line. At the second lead time when predictability has dropped to moderate levels (middle row), the scatter plots show a weakened dependency structure between ensemble members and between member one and the true state. At the third lead time when all predictability has been lost (bottom row), the ensemble mean in the time series shows only random fluctuations from the climatological mean (dashed line) due to finite ensemble size. The conditional dependence between the members and the true state has completely disappeared, as evidenced by the circular cloud of points in the two scatter plots. Note that whereas the climatological variance in each ensemble member is the same under all three predictability regimes (quantified by the climatological variance $\sigma_x^2=1$), the climatological variance of the ensemble mean decreases as one transitions from high to no potential predictability.

\section{Insufficiency of Spread-Error Diagnostics}\label{sec:pitfalls}
To better understand the underlying structure of the Spread-Error difference and Reliability Budget defined in Section \ref{sec:metrics}, both expressions are considered in statistical expectation when no assumption is made regarding exchangeability between the ensemble members and true state.

When exchangeability is only assumed between ensemble members themselves (i.e., not with truth), the expected value of the ensemble variance is given by \citepalias{marzban2011effect}:
\begin{equation}\label{eq:ens_spread_expect}
        \mathbb{E}[S_{e_t}^2]=\mathbb{E}[\langle S_{e_t}^2\rangle]\equiv \sigma_e^2=\sigma_x^2(1-r).
\end{equation}
A derivation of $\mathbb{E}[S_{e_t}^2]$ may be found in \citetalias{dirkson2025impact}, and the extension to the mean ensemble variance over $T$ forecasts follows trivially. Equation \ref{eq:ens_spread_expect} states that $\sigma_e^2$, which we define as the mean ensemble spread under expectation, is a fraction of the forecast climatological variance, $\sigma_x^2$, where the fraction is determined by $0\leq1-r\leq1$. Recall that $r$ is the inter-member correlation, such that $\sigma_e^2$ is small when $r$ is close to one (reflecting high potential predictability), and $\sigma_e^2$ is large (and close to $\sigma_x^2$) when $r$ is close to zero (reflecting low potential predictability). The ratio $\sigma_e^2/\sigma_x^2$ has been widely used to estimate potential predictability under perfect model assumptions \citep[e.g.][]{murphy1988impact,rowell1998assessing,boer2004long}, which can be seen here to only depend on the inter-member correlation $r$.

\citetalias{dirkson2025impact} showed that when ensemble members are exchangeable with one another, the MSE of the ensemble mean relative to the true state (Eq. \ref{eq:mse_tau}) has expectation:
\begin{equation}\label{eq:mse_tau_expect}
\begin{aligned}    
    \mathbb{E}[\mathrm{MSE}_{\tau}] =\frac{\sigma_e^2}{n} + \sigma_y^2 + \mathrm{Cov}(X_i,X_j) - 2 \mathrm{Cov}(X_i,Y) + \left(\mu_x - \mu_y \right)^2.
\end{aligned}
\end{equation}
The first term in Eq. \ref{eq:mse_tau_expect}, $\sigma_e^2/n$, represents the expected ensemble spread (as defined in Eq. \ref{eq:ens_spread_expect}) divided by the ensemble size, and reflects sampling variability in the ensemble mean due to finite ensemble size. This term tends to zero as $n\to\infty$. The remaining terms represent, respectively, the climatological variance of the truth, the inter-member covariance, twice the negative of the member-truth covariance, and the squared climatological mean bias.

Substituting Eqs. \ref{eq:ens_spread_expect} and \ref{eq:mse_tau_expect} into the expected value of the Spread-Error difference against the truth (Eq. \ref{eq:delta_tau}), we obtain:
\begin{equation}\label{eq:delta_tau_expect}
     \mathbb{E}[\delta_\tau]= -(\Delta\mu)^2 + \Delta\sigma^2 - 2\Delta\Sigma,
\end{equation}
where:
\begin{subequations}
\begin{align}
\Delta \mu~  & = \mu_x - \mu_y, \label{eq:delta_mu} \\
\Delta \sigma^2 & = \sigma_x^2 - \sigma_y^2, \label{eq:delta_sigma} \\
\Delta \Sigma~  & = \mathrm{Cov}(X_i,X_j) - \mathrm{Cov}(X_i,Y), \label{eq:delta_cov}
\end{align}
\end{subequations}
are respectively the climatological mean bias, the climatological variance bias, and the inter-member covariance bias. Note that the expected value of $\delta_{\hat{\tau}}$ (i.e., the Spread-Error difference based on observations given by Eq. \ref{eq:delta_tau_hat}, but with the observation error variance subtracted) is equivalent to Eq. \ref{eq:delta_tau_expect}, assuming the observation error variance is correctly specified. This expression reveals that all three conditions for reliability from Section \ref{sec:framework} are in fact embedded within the Spread-Error difference under expectation, expressed as biases. When all three conditions for reliability up to second order are satisfied, such that $\Delta \mu=\Delta \sigma^2=\Delta \Sigma=0$, it follows that $\mathbb{E}[\delta_\tau]=0$ as well. That is, when ensemble members are indeed exchangeable with the truth, the Spread-Error relationship is satisfied, as previously concluded in \cite{fortin2014should} and \citetalias{dirkson2025impact}.

As the Reliability Budget of \citetalias{rodwell2016reliability} (given by Eq. \ref{eq:reliability_budget}) explicitly removes the impact of mean bias on the Spread-Error difference, its expected value can be seen as a special case of Eq. \ref{eq:delta_tau_expect} above but with $\Delta\mu=0$:
\begin{equation}\label{eq:reliability_budget_expect}
    \mathbb{E}[\delta_{\hat{\tau}}^\nu] = \Delta\sigma^2 - 2\Delta\Sigma.
\end{equation}
As stated earlier, \citetalias{rodwell2016reliability} interpreted $\delta_{\hat{\tau}}^\nu=0$ as indicating that the ensemble spread is perfect, assuming the observation error variance subtracted in the original budget is correctly specified. Equation \ref{eq:reliability_budget_expect} can also be understood by considering the expected ensemble spread  (Eq. \ref{eq:ens_spread_expect}), which can be written $\sigma_e^2=\sigma_x^2-\mathrm{Cov}(X_i,X_j)$. Clearly, the relevant conditions for reliability that pertain to the ensemble spread are conditions 2 and 3, which respectively require that $\sigma_x^2=\sigma_y^2$ and $\mathrm{Cov}(X_i,X_j)=\mathrm{Cov}(X_i,Y)$. When these two conditions are satisfied, it is indeed the case that $\Delta \sigma^2=\Delta \Sigma=0$, such that $\mathbb{E}[\delta_{\hat{\tau}}^\nu]=0$.

It is critical to recognize, however, that while the Spread-Error difference and Reliability Budget are respectively capable of diagnosing perfect second-order reliability and perfect ensemble spread, neither is sufficient in diagnosing these conditions. In the case of the Spread-Error difference under expectation (Eq. \ref{eq:delta_tau_expect}), $\mathbb{E}[\delta_\tau]=0$ only requires a balance among three distinct biases. Because this solution constrains the sum of terms $-(\Delta\mu)^2$, $\Delta\sigma^2$, and $-2\Delta\Sigma$ to be zero, there exist infinitely many solutions to $\mathbb{E}[\delta_\tau]=0$ that do not correspond to the reliability conditions $\Delta \mu=\Delta \sigma^2=\Delta \Sigma=0$. To show reliability up to second order with the Spread-Error relationship, two out of the three terms in Eq. \ref{eq:delta_tau_expect} must also be shown to be zero. One such possibility, which aligns with \cite{johnson2009reliability}, is to show that mean bias is zero and that the forecast climatological variance equals that of the truth (reliability conditions 1 and 2 in Section \ref{sec:framework}). Similarly, the Reliability Budget is satisfied when $\Delta\sigma^2=2\Delta\Sigma$, without requiring each term to be zero. Hence, $\mathbb{E}[\delta_{\hat{\tau}}^\nu]=0$ is necessary but insufficient for diagnosing perfect ensemble spread. To do so, one must also show that climatological variance equals that of the truth (reliability condition 2 in Section \ref{sec:framework}). The solution $\Delta\sigma^2=2\Delta\Sigma$ can also be seen as a special case of the Spread-Error difference under expectation when forecasts do not contain mean bias.

\section{Idealized Ensemble System}\label{sec:idealized}
We now expand upon these theoretical results using the idealized system from Section \ref{sec:framework} based on the MVN distribution (as in \citetalias{marzban2011effect} and \citetalias{wilks2011reliability}). Because the MVN distribution is fully defined up to second order, it is a natural candidate to study both diagnostics. Results are placed into further context by considering the remaining diagnostics described in Section \ref{sec:metrics}. In all of the idealized experiments considered throughout the remainder of this study, a sample size of $T = 10^5$ is used to eliminate nearly all sampling uncertainty in each diagnostic due to finite $T$. A relatively large ensemble size of $n = 50$ is furthermore used to essentially eliminate the sampling errors that otherwise affect the reliability diagram and the reliability component of the CRPS for small ensembles. Finally, the climatological distribution for the true state is set to the standard normal distribution ($\mu_y = 0$ and $\sigma_y^2 = 1$). 

When presenting the reliability diagram in this section and the next, we consider three separate events corresponding to exceeding specific quantiles of the true climatological distribution, denoted $q_p$, where $p\in(0,1)$ is a probability. The quantiles correspond to probabilities $p=0.5$ (the climatological median and mean), $p=2/3$ (the upper tercile, often used to define above-normal events in seasonal forecasting), and $p=0.95$ (a moderately extreme event occurring in fewer than 5\% of cases). For reference, these are equal to $q_{0.5}=0$, $q_{0.6\overline{6}}\approx0.43$, and $q_{0.95}\approx1.64$ for the standard normal distribution.

\begin{figure*}[t]
\centering
\noindent\includegraphics[width=1.0\textwidth,angle=0]{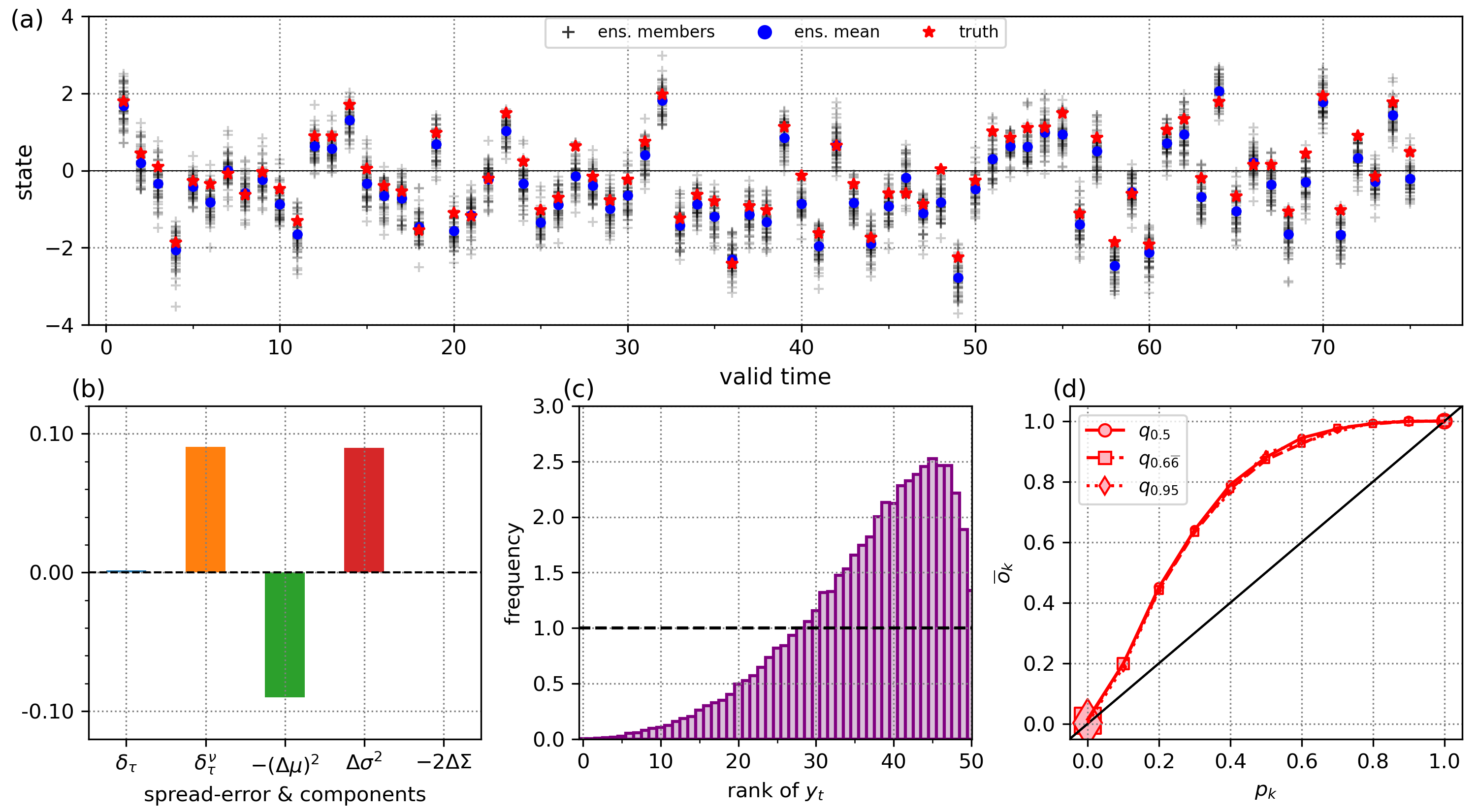}\
\caption{Ensemble forecasts simulated with a multivariate normal distribution subject to a climatological mean bias and climatological variance excess: Exp1 in Table \ref{tab:t1}. (a) Time series of the first 75 samples of the ensemble members, ensemble mean, and true state; (b) Spread-Error difference ($\delta_\tau$; Eq. \ref{eq:delta_tau}), Reliability Budget ($\delta_\tau^\nu$; Eq. \ref{eq:reliability_budget}), and the three components of the Spread-Error decomposition in expectation (Eq. \ref{eq:delta_tau_expect}); (c) rank histogram; (d) reliability diagram for three exceedance events based on quantiles of the true climatological distribution. Diagnostics in (c)-(d) were computed using all $T=10^5$ samples.}\label{f2}
\end{figure*}

\subsection{Case Studies}\label{sec:case_studies}
Table \ref{tab:t1} summarizes the MVN parameters for three case studies: Exp1, Exp2, and Exp3. Each experiment is designed to better understand the solutions to $\mathbb{E}[\delta_\tau]=0$ and $\mathbb{E}[\delta_\tau^\nu]=0$ that are inconsistent with exchangeability. In all three experiments, $R=0.9$ is fixed to represent high actual predictability.
\begin{table}[ht] 
\centering 
\begin{tabular}{lccc} 
\hline 
\textbf{Parameter} & \textbf{Exp1} & \textbf{Exp2} & \textbf{Exp3} \\ \hline $\mu_x$ & $-0.30$ & $0.00$ & $0.00$ \\ 
$\mu_y$ & $0.00$ & $0.00$ & $0.00$ \\ 
$\sigma_x^2$ & $1.09$ & $4.00$ & $0.60$ \\ 
$\sigma_y^2$ & $1.00$ & $1.00$ & $1.00$ \\ 
$\mathrm{Cov}(X_i,X_j)$ ($r$) & $0.94$ ($0.86$) & $3.30$ ($0.83$) & $0.50$ ($0.83$) \\ 
$\mathrm{Cov}(X_i,Y)$ ($R$) & $0.94$ ($0.90$) & $1.80$ ($0.90$) & $0.70$ ($0.90$) \\ 
\hline 
\end{tabular} 
\caption{Multivariate normal distribution parameters for experiments Exp1, Exp2, and Exp3 (rounded to two decimal places). Values in parentheses denote correlation coefficients.} \label{tab:t1} 
\end{table}

Ensemble forecasts in Exp1 are subject to both a negative mean bias ($\Delta\mu<0$) and excess climatological variance ($\Delta\sigma^2>0$). With $\Delta\Sigma=0$, we have that $r<R$, meaning that actual predictability exceeds potential predictability. Such a case could arise in real ensemble forecasts if ensemble spread were inflated to account for the presence of mean bias. Time series of the simulated forecasts show the negative bias (Fig. \ref{f2}a), with the truth systematically in the upper portion of the ensemble. However, due to the balance between the square of the mean bias and climatological variance bias, the Spread-Error relationship is nonetheless satisfied ($\delta_\tau\approx0$; Fig. \ref{f2}b), consistent with the theoretical result presented in Eq. \ref{eq:delta_tau_expect}. As the Reliability Budget removes the impact of mean bias from the Spread-Error relationship, it diagnoses overdispersion ($\delta_\tau^\nu>0$). Similarly, the rank histogram (Fig. \ref{f2}c) shows overdispersion (dome-shaped distribution) and bias (sloped distribution). The curves on the reliability diagram (Fig. \ref{f2}d) are consistent with underforecasting, implying that the negative bias dominates. In the absence of the bias (not shown), the reliability curves are S-shaped and indicative of underconfidence under the same covariance structure. Exp1 reemphasizes a known result that direct comparison of ``Spread'' and ``Error'' implicitly assumes that forecasts are unbiased.

\begin{figure*}[t]
\centering
\noindent\includegraphics[width=1.0\textwidth,angle=0]{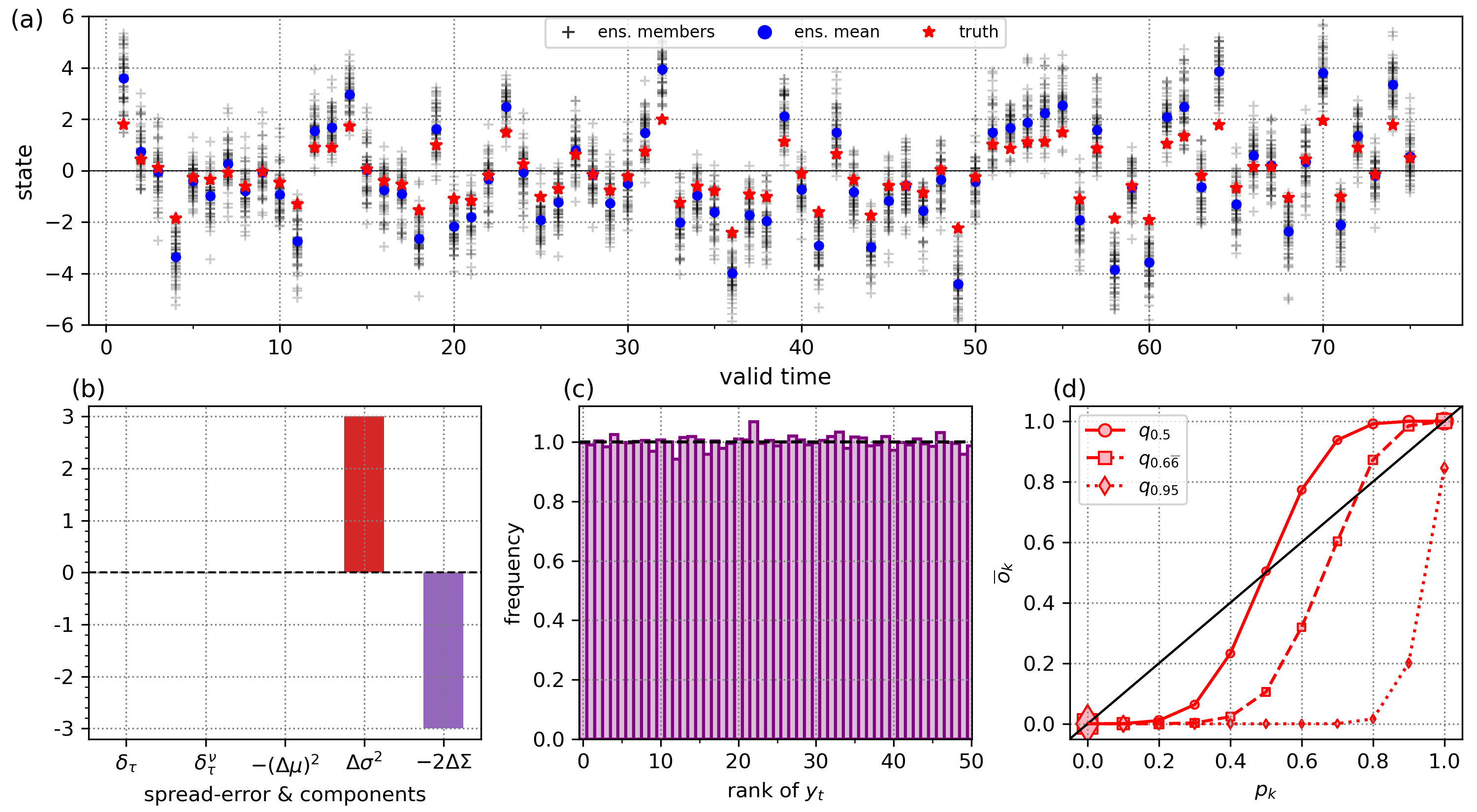}\
\caption{As in Fig. \ref{f2}, but for Exp2: climatological variance excess balanced by an inter-member covariance bias twice its magnitude.}\label{f3}
\end{figure*}

\begin{figure*}[t]
\centering
\noindent\includegraphics[width=1.0\textwidth,angle=0]{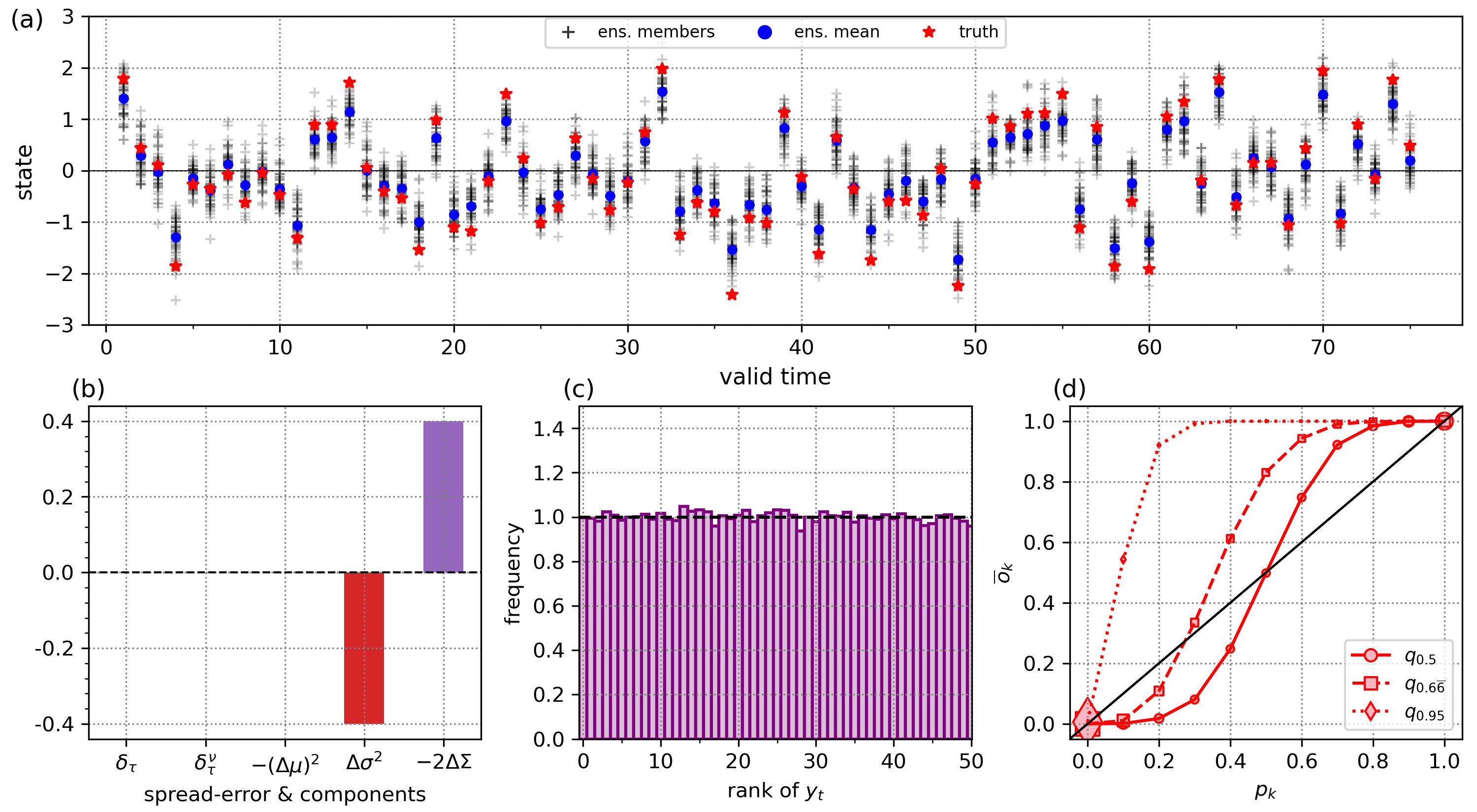}\
\caption{As in Fig. \ref{f2}, but for Exp3: climatological variance deficiency balanced by an inter-member covariance bias twice its magnitude.}\label{f4}
\end{figure*}

In Exp2, ensemble forecasts are unbiased ($\Delta\mu=0$) but have excess climatological variability ($\Delta\sigma^2>0$). The inter-member covariance bias is set so that $\Delta\sigma^2=2\Delta\Sigma$. We again have $r<R$, although the outcomes $r=R$ and $r>R$ are also possible under the same two constraints. Under this joint-distribution structure, a notable feature in the time series of Fig. \ref{f3}a is that the true state is found to typically reside among the least anomalous members (i.e., those closest to the climatological mean $\mu_x=\mu_y=0$) whenever the forecast ensemble mean is sufficiently different from the climatological mean. This non-random structure implies that the truth is not statistically indistinguishable from each ensemble member at any particular validation time. Despite this behavior, the Spread-Error difference and Reliability Budget are both essentially zero (Fig. \ref{f3}b). 

Interestingly, the rank histogram is also found to be flat for Exp2 (Fig. \ref{f3}c), apart from small random variations from finite $T$. Recall that a flat rank histogram means that the true state falls between each ordered pair of ensemble members an equal number of times across all forecasts, and is usually interpreted as meaning perfect ensemble dispersion. Roughly speaking, the flat distribution implies that the structural departures from reliability evident in the time series must balance over the full set of forecasts. For example, the too few number of times the truth falls between the two largest members when $\overline{x}_t>>0$ must be balanced by an excess number of times the truth falls between the two largest members under different forecast scenarios. \cite{bishop2008bayesian} also showed that a flat rank histogram can be produced under excessive climatological variance, and demonstrated that by stratifying the forecasts based on the extremity of the ensemble mean, unreliability then becomes immediately apparent.

The reliability diagram for Exp2 (Fig. \ref{f3}d) shows clear departures from reliability. Exceeding the median quantile is forecast under-confidently (S-shaped curve), which is found to be a result of actual predictability exceeding potential predictability ($R>r$); this will become clearer in the following section. The events corresponding to the exceedance of quantiles $q_{0.6\overline{6}}$ and $q_{0.95}$ are generally over-forecast, with over-forecasting being substantially more pronounced for the extreme event. This tendency for the two larger quantiles aligns with the behavior observed in the time series, where the ensemble members are substantially more extreme than the true state whenever the ensemble mean is also extreme ($\overline{x}_t>>0$).  

In Exp3 (Fig. \ref{f4}), the condition $\Delta\sigma^2=2\Delta\Sigma$ is again satisfied, but this time the climatological variance is underestimated ($\Delta\sigma^2<0$). In this scenario, the true state resides consistently among the most anomalous (extreme) ensemble members when the ensemble mean is sufficiently different from climatology (Fig. \ref{f4}a). Still, the Spread-Error difference, Reliability Budget, and rank histogram all suggest perfect reliability (Fig. \ref{f4}b,c). Like for Exp2, the reliability diagram exposes the conditional departures from reliability (Fig. \ref{f4}d). In this case, the above-normal and extreme events are substantially under-forecast (due to insufficient climatological variance), whereas exceeding the median quantile is forecast under-confidently (since $R>r$).

\subsection{Generality of non-detection}\label{sec:general_results}
\begin{figure*}[t]
\centering
\noindent\includegraphics[width=1.0\textwidth,angle=0]{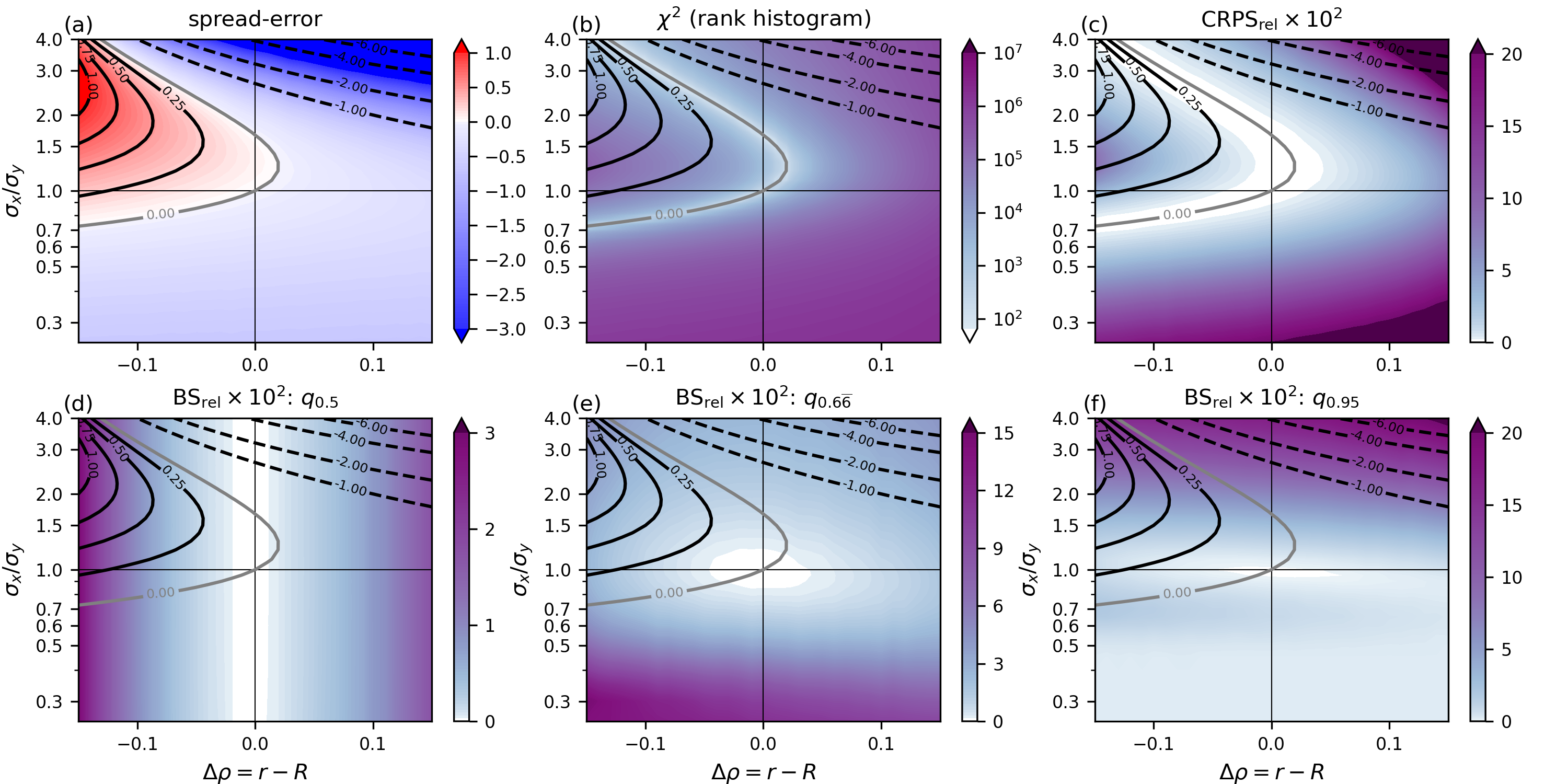}\
\caption{Reliability diagnostics as a function of the climatological standard deviation ratio $\sigma_x/\sigma_y$ and linear predictability bias $\Delta\rho$. Contours are constant values of $\Delta\sigma^2-2\Delta\Sigma$ (i.e., the Reliability Budget under expectation). The only coordinate corresponding to actual reliability is $\sigma_x/\sigma_y=1$ and $\Delta\rho=0$.}\label{f5}
\end{figure*}

To investigate the generality of the results from Exp2 and Exp3, we now consider a range of MVN distributions covering a broad parameter space. We also extend the list of diagnostics to include the reliability component of the CRPS, $\mathrm{CRPS_{rel}}$. Here, the climatological variance bias is expressed as a ratio of standard deviations, $\sigma_x/\sigma_y$, which is varied from 0.25 to 4. We then define the linear predictability bias as:
\begin{equation}\label{eq:delta_rho}
    \Delta\rho = r - R,
\end{equation}
which is varied from -0.15 to 0.15 under a moderate-to-high predictability regime of $R=0.8$. Forecasts are unbiased ($\Delta\mu=0$) to focus on cases where marginal calibration is not satisfied due to a climatological variance bias. A grid of MVN parameters spanning this space is constructed and an idealized experiment is carried out for each combination of parameters. 

Spread-Error differences (computed with Eq. \ref{eq:delta_tau}) are shown in Fig. \ref{f5}a as a function of $\Delta\rho$ and $\sigma_x/\sigma_y$. Contours are overlaid corresponding to constant values of $\Delta\sigma^2-2\Delta\Sigma$ (i.e., the expected value of the Reliability Budget) obtained directly from the parameters of the MVN distribution. The Spread-Error differences align indistinguishably from those contours, as anticipated since $\Delta\mu=0$ and $T$ is large. While the only coordinate corresponding to actual exchangeability between ensemble members and the truth is $\sigma_x/\sigma_y=1$ and $\Delta\rho=0$, the Spread-Error difference is (approximately) zero for all experiments satisfying $\Delta\sigma^2-2\Delta\Sigma=0$ (i.e., along the full extent of the gray contour), which again is consistent with theory. Similarly though, the $\chi^2$ statistic, which quantifies deviations from uniformity of the rank histogram, is found to be minimized (indicating that non-uniformity is unlikely) along the full extent of the same contour (Fig. \ref{f5}b). Adding to this result, the $\mathrm{CRPS_{rel}}$ term is also approximately zero under the same covariance structure (Fig. \ref{f5}c). Diagrams like those in Fig. \ref{f5} have also been constructed for other values $R$ and the same conclusions hold (not shown).

Given these experimental results for the Spread-Error relationship, rank histogram, and $\mathrm{CRPS_{rel}}$, as well as the theoretical results presented earlier, we arrive at the following proposition:
\begin{proposition}
\label{prop1}
For the sequence of random variables $Y,X_1,\dotsc,X_n$, ensemble members $X_1,\dotsc,X_n$ are {\bf unconditionally reliable} up to second order if the following conditions hold: \\ 1. $\Delta\mu=0$ \\ 2. $\Delta\sigma^2=2\Delta\Sigma$. 
\end{proposition}
\noindent Unless it is also the case that $\Delta\sigma^2=2\Delta\Sigma=0$, {\it unconditional reliability} represents a non-ideal form of reliability, as it implies that the truth only behaves like an ensemble member up to second order when considered unconditionally over the full set of forecasts. The requirements for unconditional reliability (which are met along the full gray contour in each panel of Fig. \ref{f5}) are substantially weaker than those for second-order exchangeability, as unconditional reliability can be maintained even for large climatological variance biases. It is understood from Exp2 and Exp3 that, under such climatological variance biases, forecast uncertainty is poorly represented when conditioned on the forecast state itself. 

That unconditional reliability carries little significance in the presence of a climatological variance bias is further supported by the reliability diagram, summarized here by the reliability component of the Brier score, $\mathrm{BS_{rel}}$ (Fig. \ref{f5}d-f). The reliability curves for the three exceedance events only mutually follow the 1:1 line ($\mathrm{BS_{rel}}\approx0$) when $\sigma_x/\sigma_y=1$ and $\Delta\rho=0$ are both satisfied. Interestingly, the quality of forecast probabilities for exceeding the median threshold (Fig. \ref{f5}d) only depends on the correlation difference $\Delta\rho=r-R$ and not the climatological variance bias. Consequently, reliability curves follow the 1:1 line whenever $\Delta\rho=0$ is satisfied. This reflects the fact that the sign of the anomaly of the true state, conditioned on a certain fraction of ensemble members sharing the same signed anomaly, is invariant to any multiplicative scaling of the ensemble member anomalies. On the other hand, climatological variance biases can substantially degrade the quality of forecast probabilities ($\mathrm{BS_{rel}}>>0$) for the above-normal event ($q_{0.6\overline{6}}$; Fig. \ref{f5}e) and the extreme event ($q_{0.95}$; Fig. \ref{f5}f). For these events, forecast probabilities only match observed frequencies when $\sigma_x/\sigma_y=1$ and $\Delta\rho=0$ are both satisfied. In instances where $\sigma_x/\sigma_y\neq1$ and other diagnostics show unconditional reliability (along the gray contour), forecast probabilities for these events can deviate strongly from the observed frequencies. 

\section{Real Ensemble Forecasts}\label{sec:real_forecasts}
\begin{figure*}[t]
\centering
\noindent\includegraphics[width=1.0\textwidth,angle=0]{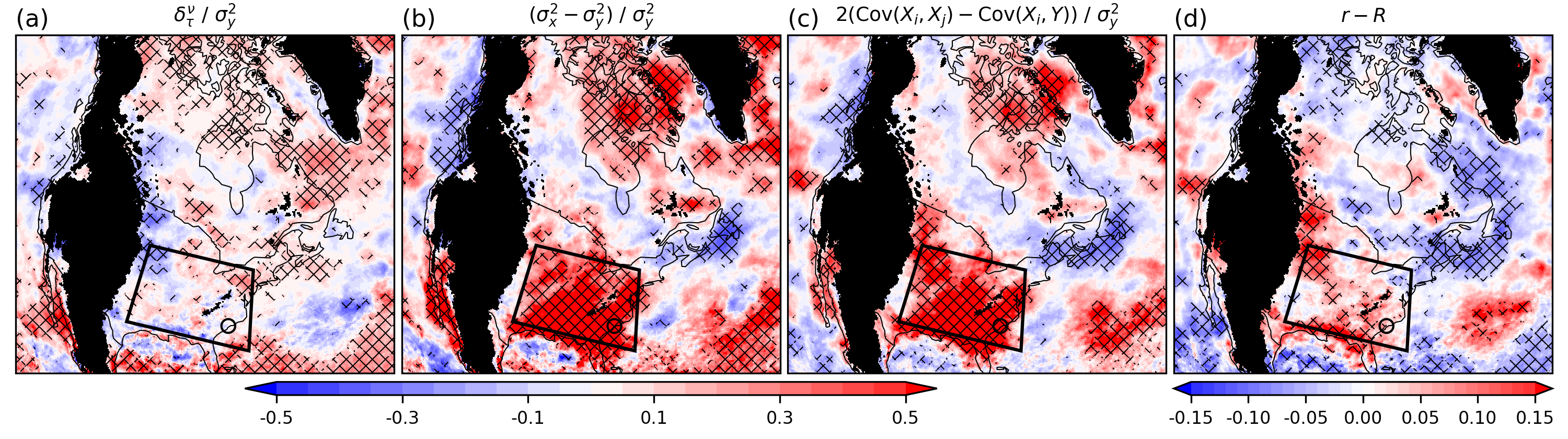}\
\caption{Statistics for 5-day GEPS temperature forecasts at 925 hPa against GDPS analysis over a two-month summer period. (a) The Reliability Budget, (b) the climatological variance bias, (c) twice the inter-member covariance bias, and (d) the linear predictability bias (potential minus actual). Quantities in panels (a)-(c) are normalized by the climatological variance of the analysis. Hatched areas indicate where the corresponding statistic differs significantly from zero at the 95\% confidence level based on bootstrapping. The black mask over mountainous regions and Greenland corresponds to locations where 925 hPa is not physically attainable in all forecasts (surface pressure $< 925$ hPa at least once). The black rectangle is discussed in the main text, and within it the black circle marks the location relevant to Fig. \ref{f7}a.}\label{f6}
\end{figure*}
\begin{figure*}[t]
\centering
\noindent\includegraphics[width=1.0\textwidth,angle=0]{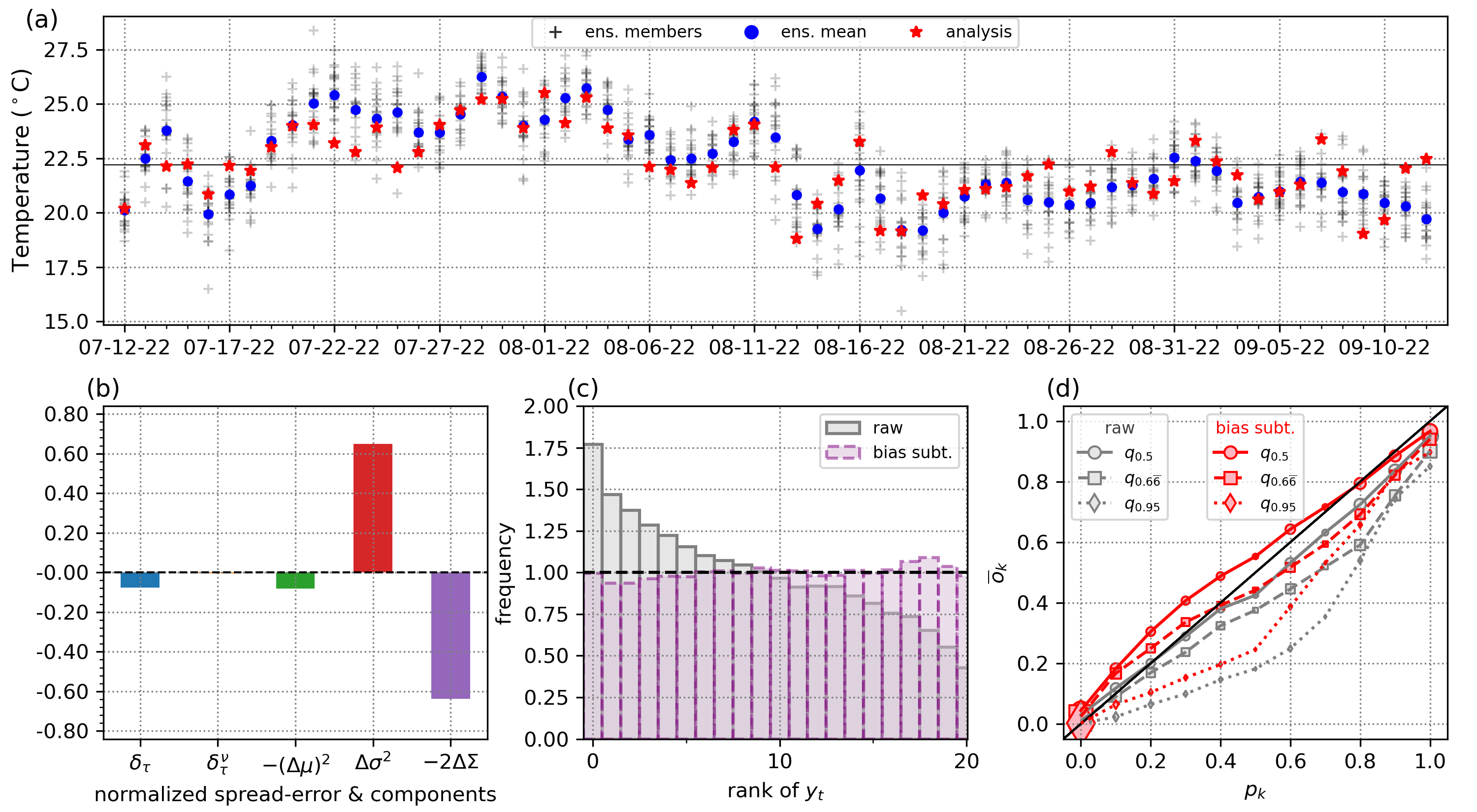}\
\caption{(a) Time series of bias-corrected forecasts at the location marked by the circle in Fig. \ref{f6}. (b)-(d) Reliability diagnostics for the forecasts in Fig. \ref{f6} inside the rectangle over the southeastern US. Quantities in (b) have been normalized by the climatological variance of the analysis. In (c) and (d), ``raw'' denotes original forecasts whereas ``bias subt.'' denotes that mean bias has been subtracted from forecasts prior to computing the diagnostic.}\label{f7}
\end{figure*}

A reasonable question following the theoretical and experimental results presented thus far is whether the highlighted covariance structure ($\Delta\sigma^2=2\Delta\Sigma$ and $\Delta\sigma^2\neq0$) that leads to unconditional reliability should ever be expected in real ensemble forecasts. This question is addressed through an example using Environment and Climate Change Canada's (ECCC) operational Global Ensemble Prediction System (GEPS), which has 20 ensemble members and is run at an approximately 25 km horizontal resolution \citep{ECCC2024}. We consider medium-range forecasts of temperature at 925 hPa valid at 00 UTC from July 12 to September 12 over a North American domain. For the reference ``truth,'' we use the analysis of ECCC's Global Deterministic Prediction System. A 5-day lead time is used to mitigate the effect of errors in the analysis itself, an assumption that rests on the analysis error being negligible relative to the forecast error. This assumption is likely reasonable over the primary area of interest for this domain discussed below, due to the relatively high density of assimilated observations constraining temperature.

Figure \ref{f6}a shows the Reliability Budget (calculated with Eq. \ref{eq:reliability_budget} but with $\sigma_o^2=0$), which is used over the traditional Spread-Error relationship to (according to \citetalias{rodwell2016reliability}) isolate deviations from reliability attributable to ``imperfect ensemble spread''---i.e., those not associated with mean bias. The two terms comprising the Reliability Budget in expectation (Eq. \ref{eq:reliability_budget_expect}), $\Delta\sigma^2$ and $2\Delta\Sigma$, are computed using sample statistics. Those terms are then normalized by the climatological variance of the analysis to fairly compare magnitudes across space. Statistical significance is assessed using bootstrap resampling with 1000 samples; 95\% confidence intervals are taken as the 2.5th--97.5th percentiles of the bootstrap distribution. 

Firstly, climatological variance biases (Fig. \ref{f6}b) and inter-member covariance biases (Fig. \ref{f6}c) show similar spatial patterns and are generally of the same sign. This behavior is also observed over other regions, atmospheric levels, and variables (not shown).  A clear explanation for why this pattern exists is beyond the scope of the present study; however, it is hypothesized to be related to initial-value predictability which should constrain $\mathrm{Cov}(X_i,X_j)$ to a magnitude similar to $\sigma_x^2$ at the analysis time. A large region in the southeastern US is highlighted (black rectangles) where the forecast climatological variance is usually substantially larger than the analysis variance, but where the terms $\Delta\sigma^2$ and $2\Delta\Sigma$ are approximately equal in magnitude. As a result of this balance, the Reliability Budget (Fig. \ref{f6}a) is statistically consistent with zero. Direct interpretation of $\delta_{\hat{\tau}}^\nu$ would lead one to falsely conclude that ensemble spread is essentially perfect in that region. Note that this behavior also occurs over other areas, such as off the coast of British Columbia and regions of Canada west of Greenland. 

To see how this result aligns with other previous findings based on the idealized experiments, time series of forecasts (after having removed mean bias) and analyses for a single location representative of this region are produced in Fig. \ref{f7}a. Like in Exp2 from before, which also had a positive climatological variance bias and $\delta_\tau^\nu\approx0$, the analysis often falls among the least anomalous members when the ensemble mean is different from climatology. The average Spread-Error difference over the full region is negative (Fig. \ref{f7}b) as a result of a net positive mean bias, whereas the average Reliability Budget is approximately zero (as expected). The rank histogram is sloped as a result of the positive bias, but is approximately flat if the mean bias at each grid point is removed prior to its construction (Fig. \ref{f7}c). The reliability diagram (Fig. \ref{f7}d) shows consistent over-forecasting for the original forecasts due to the positive mean bias. Even if mean bias is removed, however, exceeding $q_{0.95}$ is still over-forecast, indicating that the positive climatological variance bias dominates this event (as in Exp2). The reliability curves for $q_{0.5}$ and $q_{0.6\overline{6}}$ are reverse S-shaped, likely because potential predictability mostly exceeds actual predictability in this region ($r>R$ in Fig. \ref{f6}d).

\section{Qualifying reliability in terms of spread}\label{sec:qualifying}
We now focus more specifically on the issue with framing reliability up to second order in terms of spread or dispersion (i.e., perfect dispersion, underdispersion, overdispersion) when assessed over a collection of forecasts. As outlined in Section \ref{sec:intro}, this framing is especially common in ensemble data assimilation and medium-range forecasting, where consistency between the MSE of the ensemble mean and the mean ensemble spread is often interpreted as evidence of ``perfect spread.'' However, as shown in the preceding sections, it is possible to satisfy Spread-Error equality without satisfying reliability conditions 1-3 (even in the absence of unconditional bias). The same limitation applies to the rank histogram under the assumption of joint normality, a diagnostic which is also often used to infer perfect ensemble dispersion. These findings indicate that the terms ``perfect spread'' and ``perfect dispersion'' are ill-defined through Spread-Error or rank-based framings.

A natural extension to this topic pertains to whether the classifications of ``underdispersion'' and ``overdispersion'' remain valid through these framings when climatological variance is misrepresented. This labeling is important not only for accurately describing ensemble forecast behavior, but also for guiding efforts to improve the representation of forecast uncertainty in operational systems through inflationary techniques (either statistical or physical). Such methods, applied during data assimilation or forecast integration, are typically motivated by the assumption that if ``Spread'' is less than ``Error,'' forecasts must be underdispersive and ensemble spread should be increased \citep[e.g.,][]{mitchell2000adaptive,bowler2017inflation,mctaggart2022using,inverarity2023met}.

\subsection{Case Study}\label{sec:qualifying_casestudy}
\begin{figure*}[t]
\centering
\noindent\includegraphics[width=1.0\textwidth,angle=0]{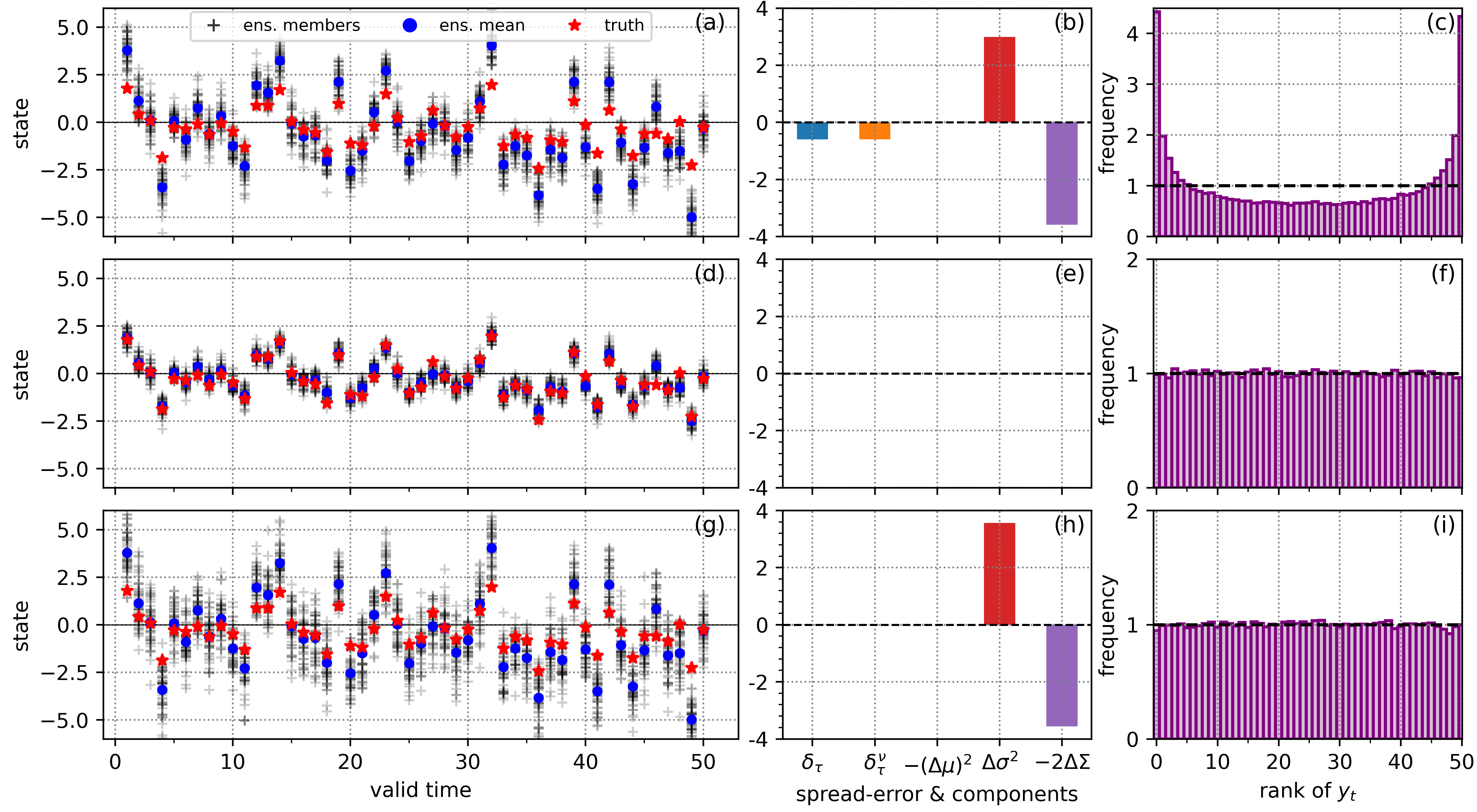}\
\caption{Case study from \citetalias{marzban2011effect} and \citetalias{wilks2011reliability} with a climatological variance bias ($\sigma_x/\sigma_y=2$) and high potential and actual predictability ($r=R=0.9$). Top row: original forecasts; middle row: after calibrating with quantile mapping; bottom row: after applying additive inflation following a misdiagnosis of underdispersion. Only the forecasts in (d)-(f) are truly reliable, even though the forecasts in (g)-(i) achieve Spread-Error equality and a flat rank histogram.}\label{f8}
\end{figure*}
To introduce this problem, we revisit a case study considered in \citetalias{marzban2011effect} and \citetalias{wilks2011reliability} also generated with MVN distribution. In that experiment, forecasts were subject to a positive climatological variance bias ($\sigma_x/\sigma_y=2$) and exhibited a high level of potential predictability that matches actual predictability ($r=R=0.9$). \citetalias{marzban2011effect} found that the forecasts for this experiment produced a U-shaped rank histogram, despite the positive climatological variance bias. \citetalias{wilks2011reliability} furthermore showed that the Spread-Error relationship is consistent with an underdispersive ensemble, and then used an argument based on conditional expectation to determine that the ensemble members are too extreme (distant from the climatological mean) on average. \citetalias{wilks2011reliability} then concluded that the forecasts are therefore ``underdispersive.'' Diagnostic results in Fig. \ref{f8} are consistent with the combined findings of \citetalias{marzban2011effect} and \citetalias{wilks2011reliability}. In particular, the truth routinely falls among the least anomalous members (Fig. \ref{f8}a), and the Spread-Error difference, Reliability Budget, and rank histogram indicate that the forecasts are ``underdispersive'' following traditional classification (Fig. \ref{f8}b,c).

However, if the forecasts in this case study were truly underdispersive, this should imply that ensemble spread must increase in order for the forecasts to become perfectly reliable. To test this directly, we appeal to calibration principles from statistical postprocessing. Since the joint distribution is MVN and the linear predictability condition $r=R$ is already satisfied, the ensemble forecasts become effectively calibrated through the following transformation:
\begin{equation*}
x_{i,t}^* = G^{-1}[F_i(x_{i,t})],
\end{equation*}
where $G^{-1}$ is the inverse CDF of the true-state climatology, $F_i$ is the CDF of the $i$th member climatology (identical across members), and $x_{i,t}^*$ denotes the calibrated ensemble member at time $t$. This transformation is more commonly known as quantile mapping in statistical postprocessing \citep[e.g.][]{wood2002long} and enforces that $x_{i,t}^*$ has the same climatological distribution as the true state $y_t$. Note that in real postprocessing applications, the underlying distributions $F_i$ and $G$ are not known and must be estimated from past forecasts and observations. Here, we are not limited by such constraints and can proceed to study this question analytically. 

For the case studies from \citetalias{marzban2011effect} and \citetalias{wilks2011reliability}, the transformation above simplifies to:
\begin{equation*}
x_{i,t}^* = \frac{\sigma_y}{\sigma_x}x_{i,t},
\end{equation*}
since $F_i$ and $G$ are normal and $\mu_x=\mu_y=0$. Because $\sigma_y/\sigma_x<1$, calibration is therefore achieved through a multiplicative {\it deflation} of each ensemble member. As the adjustment to each member is multiplicative, it does not change the correlations $r$ and $R$. After calibration, it follows that the climatological variance is equal to $\sigma_y^2$ and the inter-member correlation remains equal to $r$. By referring to the expected ensemble spread given by Eq. \ref{eq:ens_spread_expect}, we can conclude that the ensemble spread is in fact reduced following calibration rather than increased. Despite ``Spread'' being less than ``Error'' and the rank histogram being U-shaped for the original forecasts, the original forecasts are therefore not ``underdispersive.'' The forecasts and diagnostics following calibration are shown in Fig. \ref{f8}d-f, and confirm that all of the conditions for reliability are satisfied ($\Delta\mu=\Delta\sigma^2=\Delta\Sigma=0$).

It is noteworthy that if the original forecasts were incorrectly interpreted as being underdispersive and one decided that spread must therefore be increased, one could do so statistically without affecting the MSE of the ensemble mean through additive or multiplicative inflation \citep[see e.g.,][]{bowler2017inflation}. However, such a transformation in this case will only ever result in Spread-Error equality and a flat rank histogram by satisfying the covariance structure described in the previous sections (i.e., $\Delta\sigma^2=2\Delta\Sigma$) corresponding to unconditional reliability, without ever resolving the original climatological variance bias that gave rise to the unreliability---in fact, the climatological variance bias only worsens following inflation. An example of this based on additive inflation \citep{mitchell2000adaptive}, whereby random Gaussian noise has been added to each ensemble member (with mean zero and variance $\sigma_a^2=\mathbb{E}[\delta_\tau]$) is shown in Fig. \ref{f8}g-i. While Spread-Error equality and a flat rank histogram are achieved, the climatological variance bias is further increased. Note that with this intervention, both potential and actual predictability are (in this case) unnecessarily reduced, but by a greater amount for the former such that $r=R$ is no longer satisfied.

\subsection{General Results}\label{sec:qualifying_general}
\begin{figure}[t]
\centering
\noindent\includegraphics[width=0.5\textwidth]{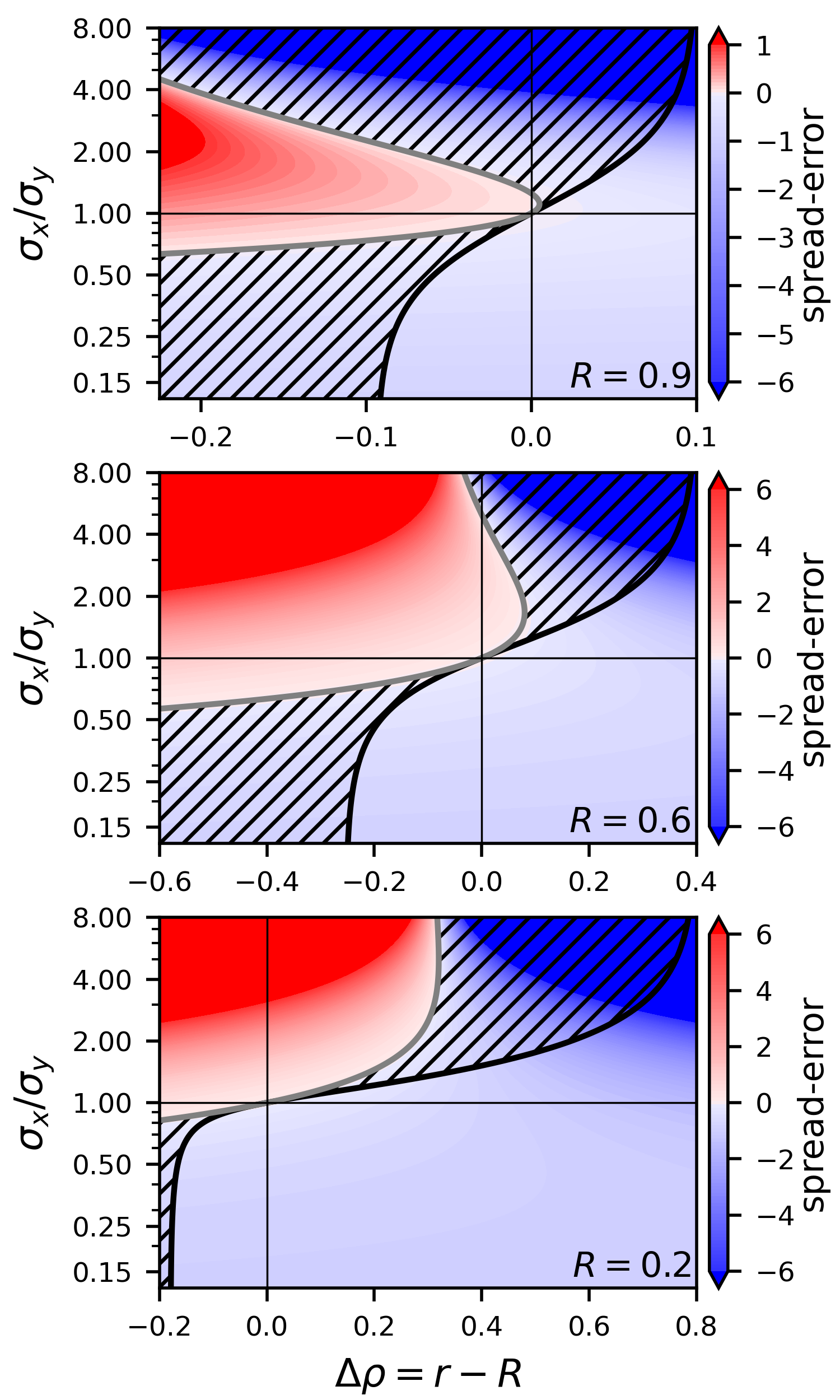}\
\caption{Spread-Error differences in expectation (Eq. \ref{eq:delta_tau_expect}) as a function of climatological standard deviations ($\sigma_x/\sigma_y$) and the linear predictability bias ($\Delta\rho$; potential minus actual) for three predictability regimes ($R$ value in each diagram). Gray contour: $\Delta\sigma^2-2\Delta\Sigma=0$ (Spread-Error equality when there is no mean bias); Black contour: $\beta^2=1$, corresponding to where ensemble spread does not change following member-by-member calibration; hatched area: where spread is reduced following calibration despite ``Spread'' being less than ``Error.''}\label{f9}
\end{figure}
To explore the topic of spread classification more generally, we consider (like in Section \ref{sec:general_results}) a range of climatological variance biases and predictability biases. We then seek to determine if the change in spread required to achieve calibration is in agreement with that indicated by Spread-Error framing.

For general calibration up to second order, we utilize a transformation known as member-by-member calibration (MBMC) \citep{doblas2005rationale,johnson2009reliability,schefzik2017ensemble}. For simplicity, reliability condition 1 is set to already be satisfied through $\mu_x=\mu_y=0$. In this situation, MBMC is given by:
\begin{equation}\label{eq:mbmc}
    x_{i,t}^* = \alpha \overline{x}_t + \beta x_{i,t}',
\end{equation}
where $x_{i,t}^*$ is the calibrated ensemble member, $x_{i,t}'=x_{i,t}-\overline{x}_t$ is the ensemble member perturbation relative to the ensemble mean, $\overline{x}_t$, and $\alpha$ and $\beta$ are scaling coefficients. These coefficients are chosen so that the calibrated forecasts jointly satisfy the climatological variance condition and the Spread-Error relationship \citep{johnson2009reliability}. As a result, MBMC implicitly also satisfies reliability condition 3 ($\Delta\Sigma=\Delta\rho=0$) through Eq. \ref{eq:delta_tau}. \cite{roberts2025unbiased} noted however that the legacy equations for $\alpha$ and $\beta$ used in \cite{doblas2005rationale} and \cite{johnson2009reliability} only satisfy these conditions for infinite-sized ensembles, and derived new coefficients for general ensemble sizes. Coefficients in the limiting case where $T\to\infty$ are derived in Appendix \ref{sec:appendix_A}, so that they may be expressed in terms of the population parameters relevant to the Spread-Error difference in expectation. 

From Eq. \ref{eq:mbmc}, the expected ensemble spread after calibration is given by $\beta^2\sigma_e^2$, such that:
\begin{itemize}
    \item if $\beta^2=1$, ensemble spread need not change,
    \item if $\beta^2>1$, ensemble spread must increase,
    \item if $\beta^2<1$, ensemble spread must decrease
\end{itemize}
to achieve calibration. Equation \ref{eq:mbmc} also tells us, however, that ensemble spread is not necessarily the only adjustment required to achieve calibration. Even in this simplified scenario where forecasts are unbiased, the ensemble mean must also be re-scaled whenever $\alpha\neq1$.    

For the results presented here, we set the climatological variance for the truth to $\sigma_y^2=1$. The ratio of climatological standard deviations $\sigma_x/\sigma_y$ is varied from 1/8 to 8 and we assume an ensemble size of $n=50$ (results change very little for smaller $n$). The linear predictability bias, $\Delta\rho$, spans different ranges for three predictability regimes: $R=0.9$ (high predictability), $R=0.6$ (moderate predictability) and $R=0.2$ (low predictability). The expected value of the Spread-Error difference (Eq. \ref{eq:delta_tau_expect}) is plotted as a function of $\sigma_x/\sigma_y$ and $\Delta\rho$ in Fig. \ref{f9}. 

Firstly, the contours $\beta^2=1$ (spread doesn't change following calibration) and $\Delta\sigma^2-2\Delta\Sigma=0$ (Spread-Error equality) only coincide when both $\sigma_x/\sigma_y=1$ and $\Delta\rho=0$, consistent with what is expected when forecasts are truly reliable. However, the fact that the contours otherwise do not match, and moreover extend into regions where these conditions are not met, implies that ``perfect spread'' is ill-defined, whether framed in terms of Spread-Error equality or in terms of the spread required to achieve calibration. For all three levels of predictability, we find that the region where ``Spread'' is larger than ``Error'' also aligns with where $\beta^2<1$ (to the left of the $\beta^2=1$ contour), meaning that spread must indeed be reduced to achieve calibration. As such, we conclude that the ``overdispersive'' classification may be considered generally accurate. With that said, it is important to recognize that a reduction in spread is not necessarily the only adjustment to the ensemble needed to achieve calibration when the Spread-Error difference is positive---that is only the case when $\alpha=1$ as well. Therefore, while the term ``overdispersion'' may accurately describe the directionality of spread change needed for reliability, it can still be an oversimplified description of how ensemble forecasts deviate from reliability up to second order.

For all three levels of predictability, the region where ``Spread'' is less than ``Error'' can be divided into two sub-regions: one where $\beta^2>1$ (to the right of the $\beta^2=1$ contour)---meaning ensemble spread must increase for calibration---and another where $\beta^2<1$---the hatched area, where ensemble spread must decrease. The classification of forecasts as underdispersive using Spread-Error framing only aligns universally with the calibration-based classification (i.e., $\beta^2>1$) in two specific situations: (1) when $\sigma_x=\sigma_y$  (no climatological variance bias), and (2) within the quadrant where both $\sigma_x/\sigma_y<1$ (climatological variance is too small) and $r>R$ (potential predictability exceeds actual predictability). In these two situations, ensemble spread must indeed increase to achieve calibration, consistent with Spread-Error framing. However, in the other quadrants where ``Spread'' is less than ``Error'' is possible, it is also possible to have $\beta^2<1$, implying that spread must actually be further reduced to achieve calibration. As such, the ``underdispersive'' classification should not be expected {\it a prior} to accurately reflect the adjustment in spread needed to achieve calibration when the climatological variance condition is not met. 

\section{Direct second-order assessment}\label{sec:new_diagnostic}
We have shown that the Spread-Error relationship, Reliability Budget, rank histogram, and the reliability component of the CRPS can all fail to detect departures from reliability when forecasts are subject to a climatological variance bias. As such, all of these metrics are necessary but insufficient for diagnosing reliability. In \cite{gneiting2007probabilistic}, the same conclusion for the rank histogram led those authors to recommend that marginal calibration be assessed separately. This same recommendation was offered by \cite{johnson2009reliability} regarding the Spread-Error relationship. The results presented here are broadly aligned with these conclusions, as a separate assessment of climatological reliability would counter any false diagnosis of reliability encountered previously. With that said, given that both ``perfect dispersion'' and ``underdispersion'' are ill-defined through Spread-Error and rank-based framings, one could argue that continued use of these diagnostics perpetuates a misguided notion that reliability up to second order can be reduced to a characterization based on spread/dispersion alone (after accounting for unconditional bias). 

To overcome the problem of insufficiency affecting the aforementioned diagnostics, as well as to emphasize that the description of (un)reliability should move beyond one centered on spread, we propose an alternative reliability diagnostic that directly evaluates the three conditions for exchangeability outlined in Section \ref{sec:framework}. Specifically, we propose evaluating the conditions for reliability one-by-one through the following sample-based differences against observations:
\begin{subequations} 
\begin{align} 
\Delta\tilde{\mu}_{\hat{\tau}} &=\tilde{\mu}_{x}- \tilde{\mu}_{y_o} \hspace{9.2em}\text{climatological mean bias}\label{eq:mean_diff_tilde} \\ 
\Delta\tilde{\sigma}_{\hat{\tau}}^2 &= \frac{T}{T-1}[ \tilde{\sigma}_x^2 - (\tilde{\sigma}_{y_o}^2 - \sigma_o^2)] \hspace{3em}\text{climatological variance bias} \label{eq:var_diff_tilde} \\ 
\Delta\tilde{\rho}_{\hat{\tau}} &= \frac{C_{xx}}{\tilde{\sigma}_x^2}-\frac{C_{xy_o}}{\tilde{\sigma}_x\sqrt{\tilde{\sigma}_{y_o}^2 - \sigma_o^2}}. \hspace{3.7em} \text{linear predictability bias} \label{eq:rho_diff_tilde} \end{align} 
\end{subequations} 
We collectively refer to these three differences as the Mean-Variance-Predictability (MVP) decomposition. Sample-based estimates of the quantities in Eqs. \ref{eq:mean_diff_tilde}-\ref{eq:rho_diff_tilde} are provided in Appendix \ref{sec:appendix_B} and are to be computed at individual locations and lead times over a validation period. 

As before, the subscript $\hat{\tau}$ attached to the $\Delta(\cdot)$ terms above signifies that the quantity converges in expectation to its truth-based counterpart. For the climatological mean bias (Eq. \ref{eq:mean_diff_tilde}), no adjustment is needed assuming unbiased observations. For the climatological variance bias (Eq. \ref{eq:var_diff_tilde}) and linear predictability bias (Eq. \ref{eq:rho_diff_tilde}), the observation error variance $\sigma_o^2$ must be subtracted from the climatological variance of the observations, $\sigma_{y_o}^2$. For the former, this assumes observation error is independent of the truth. For the latter, where $C_{xx}$ and $C_{xy_o}$ denote the average inter-member and member-observation covariances, the correction assumes independence between observation error and both the truth and the ensemble members. In Eq. \ref{eq:var_diff_tilde}, an additional scaling by $T/(T-1)$ provides an unbiased estimate of $\Delta\sigma^2$. No such correction is available for the correlations in the linear predictability bias, which have been chosen to quantify deviations from the third condition for reliability (as opposed to covariances) to frame the MVP decomposition through two conceptually distinct contributions: climatological reliability and reliability related to predictability.  

Beyond serving as a general approach for evaluating reliability, the MVP decomposition can also be used to ensure statistical and physics-based inflation methods---aimed at improving the representation of forecast uncertainty in real systems---target both $\Delta\tilde{\sigma}_{\hat{\tau}}^2\to0$ and $\Delta\tilde{\rho}_{\hat{\tau}}\to0$, as opposed to Spread-Error equality alone. The case study of \citetalias{marzban2011effect} and \citetalias{wilks2011reliability} in Section \ref{sec:qualifying_casestudy} served as an illustrative example of how forecast quality can be deteriorated if the wrong intervention is employed. In that scenario, the deterioration from additive inflation would be detected in both quantities.

\section{Summary}\label{sec:summary}
In this study, we have investigated the sufficiency of various reliability metrics. This was done by framing ensemble members and the verifying truth in terms of their joint probability distribution, reflecting the practical reality that reliability is assessed over collections of forecasts such that independence cannot be assumed. Within this framework, reliability corresponds to the exchangeability property \citep[\citetalias{wilks2011reliability};][]{broecker2011concept} and, up to second order, requires three conditions: (1) consistency of climatological means, (2) consistency of climatological variances, and (3) consistency of inter-member and member-truth correlations (which correspond to linear measures of potential and actual predictability). When the joint distribution of ensemble members and the truth is multivariate normal, these conditions fully characterize reliability \citepalias{marzban2011effect,wilks2011reliability}.

Under a particular structure of the joint distribution in which a climatological variance bias is matched by an inter-member covariance bias of twice its magnitude, the Spread-Error relationship, Reliability Budget, and, assuming joint normality, the rank histogram and the reliability component of the CRPS all fail to detect unreliability. Under this structure, forecasts appear unconditionally reliable over the full set of forecasts, but are unreliable at any given time. Unreliability is most evident in time series when the ensemble mean is sufficiently different from climatology. In such cases, the truth lies consistently among the most anomalous members when climatological variance is biased low, and among the least anomalous members when climatological variance is biased high. Unreliability is also evident with the reliability diagram, which confirms that this form of inconsistency is particularly consequential for extremes. Notably, this structure was also shown to be possible in an operational forecasting system.

We further argued that framing reliability in terms of spread/dispersion is broadly inadequate, even when unconditional bias is not present or has been accounted for. Not only is ``perfect spread'' ill-defined when assessed with Spread-Error or rank-based diagnostics, it is also ill-defined when reinterpreted as the spread that would be obtained following perfect calibration (via {\it a posteriori} transformation). Likewise, the classification of ensembles as ``underdispersive'' is also ill-defined, since, in the presence of a climatological variance bias, the ensemble spread may need to be further reduced to achieve calibration, even when ``Spread'' is smaller than ``Error'' and the rank histogram exhibits a U-shape. This conclusion parallels an earlier finding that conditional bias (which is statistically similar to climatological variance bias) can also be misinterpreted as underdispersion \citep{hamill2001interpretation}.

Taken together, these results underscored the importance of evaluating climatological reliability separately when applying traditional diagnostics, in agreement with earlier warnings from \cite{gneiting2007probabilistic} and \cite{johnson2009reliability}. At the same time, we highlighted the risk that continued reliance on these diagnostics may reinforce the misguided notion that reliability (up to second order) can be reduced to a characterization based on ensemble spread or dispersion. To address the fact that these diagnostics are necessary but insufficient for demonstrating reliability, we introduced the MVP reliability decomposition. This approach partitions departures from second-order reliability into three components: climatological mean bias, climatological variance bias, and linear predictability bias. By separating climatological and predictability-related contributions to unreliability, the MVP decomposition avoids the compensatory effects that can go undetected by traditional measures. This separation also promotes a more nuanced characterization of second-order reliability that extends beyond spread or dispersion alone.

Future work will focus on using the MVP decomposition to help guide improvements to reliability in ECCC's GEPS and to potentially compare traditional physics-based ensemble systems with those driven by artificial intelligence. We further intend to better understand the extent to which different inflationary techniques common in ensemble data assimilation are effective at improving second-order reliability holistically versus Spread-Error equality alone. Finally, it may be of interest to extend the MVP decomposition for applications where the joint distribution is not sufficiently described up to second order. 

\section*{Acknowledgements}
The authors thank William J. Merryfield, Jean-Fran\c{c}ois Caron, and Julia Velletta for their insightful comments on an earlier version of this work.

\appendix
\section*{Appendix A: Member-by-member Coefficients}\label{sec:appendix_A}
The goal of member-by-member calibration (MBMC) as applied in Section \ref{sec:qualifying_general} is to achieve reliability conditions 1-3 (see Section \ref{sec:framework}). In the simplified case where forecasts are unbiased ($\mu_x=\mu_y=0$), the transformation only requires reliability conditions 2 (climatological variance condition) and 3 (covariance or predictability condition) to be satisfied.

First, applying Eq. \ref{eq:mbmc} for arbitrary $\alpha$ and $\beta$ to raw members $x_{i,t}$, the climatological variance and MSE of the ensemble mean of the transformed forecasts in the limit as $T\to\infty$ become:
\begin{subequations}
\begin{align}
    \sigma_{x^*}^2 &= \alpha^2\sigma_{\overline{x}}^2 + \frac{n-1}{n}\beta^2\sigma_e^2 \label{eq:margvar_ensmean_mbmc} \\
    \mathrm{MSE}^* &=\alpha^2\sigma_{\overline{x}}^2+ \sigma_y^2-2\alpha\sigma_{\overline{x}}\sigma_y\overline{R}. \label{eq:mse_mbmc}
\end{align}
\end{subequations}
Here, 
\begin{equation*}
    \overline{R} = \frac{\mathrm{Cov}(\overline{X},Y)}{\sigma_{\overline{x}}\sigma_y}=\frac{\sigma_x}{\sigma_{\overline{x}}}R
\end{equation*}
is the correlation between the ensemble mean and the truth, and
\begin{equation*}
    \sigma_{\overline{x}}^2 = r\sigma_x^2 +\frac{\sigma_e^2}{n}
\end{equation*}
is the climatological variance of the ensemble mean (\citetalias{dirkson2025impact}). Equation \ref{eq:margvar_ensmean_mbmc} follows from $\mathbb{V}(x_{i,t}^*)=\alpha^2\mathbb{V}(\overline{x}_t)+\beta^2\mathbb{V}(x_{i,t}')$ (since the ensemble mean and ensemble perturbations are independent) and:
\begin{align*}
    \mathbb{V}(x_{i,t}') &= \sigma_x^2+\sigma_{\overline{x}}^2-2\mathrm{Cov}(X_i,\overline{X}) \\
    &= \sigma_x^2 -\sigma_{\overline{x}}^2 \\
    &= \frac{n-1}{n}\sigma_e^2,
\end{align*}
which itself follows from $\sigma_{\overline{x}}^2=\mathrm{Cov}(X_i,\overline{X})$ (see Appendix A in \citetalias{dirkson2025impact}) and Eq. \ref{eq:ens_spread_expect}.

Next, we enforce $\sigma_{x^*}^2=\sigma_y^2$ and $\mathrm{MSE}^*=\sigma_{e^*}^2(n+1)/n$, which also ensures reliability condition 3 is satisfied. Using $\sigma_{e^*}^2=\beta^2\sigma_e^2$ gives two equations in the unknowns $\alpha$ and $\beta$:
\begin{align*}
    0&=\alpha^2\sigma_{\overline{x}}^2 + \frac{n-1}{n}\beta^2\sigma_e^2 - \sigma_y^2 \\
    0&=\alpha^2\sigma_{\overline{x}}^2-\frac{n+1}{n}\beta^2\sigma_e^2 + \sigma_y^2 - 2\alpha\sigma_{\overline{x}}\sigma_y\overline{R},
\end{align*}
which are subtracted to express $\beta^2$ in terms of $\alpha$. Finally, substituting $\beta^2$ into the first expression produces a quadratic in $\alpha$, for which only the positive root is relevant. Solving gives:
\begin{subequations}
    \begin{align}
    \alpha &= \frac{\sigma_y}{\sigma_{\overline{x}}}\left[ \frac{(n-1)\overline{R} + \sqrt{(n-1)^2\overline{R}^2 + 4n}}{2n} \right] \label{eq:mmc_alpha} \\
        \beta^2 &= \frac{\sigma_y^2 - \alpha\ \overline{R}\sigma_{\overline{x}}\sigma_y}{\sigma_e^2}. \label{eq:mmc_beta} 
    \end{align}
\end{subequations}

\section*{Appendix B: MVP Reliability Decomposition}\label{sec:appendix_B}
The sample-based estimates of each term in the MVP decomposition given by Eqs. \ref{eq:mean_diff_tilde}-\ref{eq:rho_diff_tilde} are given by:
\begin{subequations}
\begin{align*} \tilde{\mu}_{x} &= \frac{1}{n}\sum_{i=1}^{n}\tilde{\mu}_{x_i} = \frac{1}{nT}\sum_{t=1}^{T}\sum_{i=1}^{n}x_{i,t} \\ 
\tilde{\mu}_{y_o} &= \frac{1}{T}\sum_{t=1}^{T}y_{o,t} \\ 
\tilde{\sigma}_x^2 &= \frac{1}{nT} \sum_{t=1}^{T}\sum_{i=1}^{n} (x_{i,t} - \tilde{\mu}_{x_i})^2 \\ 
\tilde{\sigma}_{y_o}^2 &= \frac{1}{T}\sum_{t=1}^{T}(y_{o,t} - \tilde{\mu}_{{y_o}})^2 \\ 
\mathrm{C}_{xx} &= \frac{2}{n(n-1)} \sum_{t=1}^{T}\sum_{i>j}(x_{i,t} - \tilde{\mu}_{x_i})(x_{j,t} - \tilde{\mu}_{x_j}) \\ 
\mathrm{C}_{x{y_o}} &= \frac{1}{n}\sum_{t=1}^{T}\sum_{i=1}^{n} (x_{i,t} - \tilde{\mu}_{x_i})(y_{o,t} - \tilde{\mu}_{{y_o}}). \\
\end{align*} 
\end{subequations}

\bibliographystyle{agsm}
\bibliography{references}  %%% Uncomment this line and comment out the ``thebibliography'' section below to use the external .bib file (using bibtex) .

\end{document}